\magnification=1200

\hsize 17truecm \vsize 23truecm

   \font\twelvec=msbm10 at 12pt
\font\sevenc=msbm10 at 9pt \font\fivec=msbm10 at 7pt

\newfam\co
\textfont\co=\twelvec \scriptfont\co=\sevenc \scriptscriptfont\co=\fivec

\def\ae{\mathop{\rm a.e.}\nolimits}
\def\arg{\mathop{\rm arg}\nolimits}
\def\Const{\mathop{\rm Const.}\nolimits}
\def\det{\mathop{\rm det}\nolimits}
\def\exp{\mathop{\rm exp}\nolimits}

\def\Id{\mathop{\rm Id}\nolimits}

\def\im{\mathop{\rm Im}\nolimits}

\def\re{\mathop{\rm Re}\nolimits}

\def\Hess{\mathop{\rm Hess}\nolimits}
\def\ker{\mathop{\rm Ker}\nolimits}
\def\lim{\mathop{\rm lim}\nolimits}

\def\ren{\mathop{\rm ren}\nolimits}
\def\supp{\mathop{\rm supp}\nolimits}

\def\sup{\mathop{\rm sup}\nolimits}
\def\inf{\mathop{\rm inf}\nolimits}

\def\Sum{\displaystyle\sum}
\def\e{\mathop{\rm \varepsilon}\nolimits}

\def\max{\mathop{\rm max}\nolimits}

\baselineskip 15pt

\centerline{\bf THERMODYNAMICAL EQUILIBRIUM OF VORTICES}

\centerline{\bf IN THE  ISOTROPIC BIDIMENSIONAL KAC ROTATOR}
\medskip

\centerline{H.EL BOUANANI and M.ROULEUX}
\medskip
\medskip

\centerline{Centre de Physique Th\'eorique and Universit\'e du Sud Toulon Var}

\centerline{CPT, Campus de Luminy, Case 907 13288 Marseille Cedex 9, France}

\centerline{\it{hicham.el-bouanani@cpt.univ-mrs.fr \& rouleux@cpt.univ-mrs.fr}}
\medskip
\noindent {\bf Abstract}: We consider here the problem of extrema for the Kac functional with long
range, ferromagnetic interaction, and vorticity conditions at infinity which make it not weakly
closed. Using a gradient-flow dynamics, we investigate local minima, showing strong analogies with
the Ginzburg-Landau functional in infinite volume.

\bigskip
\noindent {\bf 0. INTRODUCTION}.
\medskip
We consider here the XY or ``planar  rotator'' model and its corresponding continuous version, with
internal continuous symmetry group $O^+(2)$  and long range, ferromagnetic interaction. Such
interactions  were introduced by Lebowitz and Penrose in Statistical Physics as a generalization of
the celebrated Kac, Uhlenbeck and Hemmer model, accounting for liquid vapour phase transitions, and
giving mathematical foundations to Van der Waals theory.

A theorem of Dobrushin \& Shlosman [Si,p.78] asserts that at any inverse temperature $\beta$, every
infinite volume Gibbs state corresponding to a fairly general, translation invariant hamiltonian
$H$ with internal continuous symmetry group $O^+(2)$ on the 2-d lattice, is invariant (up to
conjugation of charge) under the group $O^+(2)$. This is absence of breakdown of continuous
symmetry. Later Bricmont, Fontaine and Landau discovered that this Gibbs state is unique. It
presents many interesting, and yet not fully understood features, such as a particular form of
phase transition at low temperature, which is characterized by the change of behavior in the
correlation functions. For the XY system they were described by Kosterlitz \& Thouless in term of
topological vortices.

Here we look instead at the free energy functional, which provides, through the mean field
approximation, a good approximation to the canonical Gibbs measure on the set of magnetizations
$m$, i.e. suitables averages of spins over mesoscopic regions. Moreover, vortices can be enhanced
by imposing some boundary condition on $H$ at infinity.

One usually consider the vector spin hamiltonian on ${\bf Z}^2$ as a limit of hamiltonians (the
thermodynamical limit) on finite lattices $\Lambda\to{\bf Z}^2$, with boundary conditions on
$\Lambda^c$. Then the free energy functional takes the form
$$\eqalign{
{\cal F}(m|m^c)&={1\over 4}\int_\Lambda dr\int_\Lambda dr'J(r-r')|m(r)-m(r')|^2\cr &+{1\over
2}\int_\Lambda dr\int_{\Lambda^c} dr'J(r-r')|m(r)-m(r')|^2+ \int_\Lambda dr
\bigl(f_\beta(m(r))-f_\beta(m_\beta)\bigr)\cr }\leqno(0.1)$$ where $f_\beta(m)=-{1\over
2}|m|^2+{1\over\beta}I(m)$ is the free energy for the mean field approximation, and $I(m)$ denotes
entropy function. Actually formula (0.1) holds also (formally) in the continuous case, see Appendix
C. Systems of spins valued in $\{+1,-1\}$, i.e. the scalar case, with long range interaction are
well understood, due in particular to a series of papers by Cassandro, DeMasi, Presutti and their
collaborators, who studied in great detail interfaces and equilibrium shapes.

In an attempt to generalize this theory to the XY model [El-BoRo], the authors have looked for
local minima of the free energy functional, and observed in numerical simulations, among other
things, vortex configurations on finite lattices $\Lambda$, induced by the vorticity on  the
boundary $\Lambda^c$, which are very similar to those arising in solutions of Ginzburg-Landau
equations, describing a superfluid or a supraconductor subject to a magnetic field. Related
examples include the 't Hooft-Polyakov monopole and the Skyrme model.

We will consider in this paper a continuous version of the rotator, that complies to the methods of
Functional Analysis. The minimization problem for the free energy functional (0.1) and its
continuous version in infinite volume has the $O^+(2)$ symmetry, which we break by imposing on the
magnetization $m$ a vorticity condition at infinity of the form $m(x)=m_\beta e^{in\theta}$,
$n\in{\bf N}\setminus 0$. Here $m_\beta\in]0,1[$ is the critical value for the mean field free
energy, and we work at a sub-critical temperature, i.e. $\beta>2$.

The obvious conjecture about a minimizer is that there is a radially symmetric vortex of degree
$n$, expressed in polar coordinates as $m(x)=u_n(r)e^{in\theta}$, for some nonnegative function
$u_n(r)$ with $u_n(r)\to m_\beta$ as $r\to\infty$, and $u_n(0)=0$. For $n=1$, this is called a {\it
hedgehog}.

A similar conjecture holds in the case of Ginzburg-Landau equation in a disc, where it is known
that the hedgehog $\psi_1$, uniquely defined, is stable, i.e. all the eigenvalues of the
self-adjoint second variation operator $L_{\psi_1}$,
$$(v|L_{\psi_1}v)={1\over 2} {d^2\over d\varepsilon^2}{\cal E}\bigl(\psi_1+\varepsilon v\bigr)|_{\varepsilon=0}\leqno(0.2)$$
are positive. This was obtained independently by Lieb and Loss [LiLo], and Mironescu [M].

In infinite volume, the situation is more subtle, due to the translational symmetry of the problem.
In fact, Ovchinnikov and Sigal showed (still for Ginzburg-Landau equation) that for any $n$, there
is a unique radially symmetric vortex $\psi_n$ of degree $n$, that minimizes ${\cal E}_{\ren
}(\psi)$ among all functions of the form $\psi(x)=u_n(r)e^{in\theta}$. Here ${\cal E}_{\ren }$ is a
renormalized energy functional. The self-adjoint second variation operator $L_{\psi_n}$ is
decomposed as channel operators in Fourier modes~;  the $n$-th mode is a positive operator, and has
not 0 for an eigenvalue, but 0 is an eigenvalue for the $n+1$-th mode (due to the translational
symmetry), and there are also negative eigenvalues for some higher modes. Thus we cannot conclude
in this case to linear stability of the radially symmetric vortex.

Minimizers of the free energy of the scalar Kac model, and their stability were investigated by
DeMasi, Presutti {\it et al.} In the 1-d case, they proved existence of a solution, called {\it
instanton}, unique modulo translations, of the minimization problem subject to the condition that
the magnetizations $m(x)$ tend to $\pm m_\beta$ as $x$ tend to infinity. Their result was rederived
lateron by Alberti and Bellettini [AlBe], and extended to higher dimensions, for a class of
functions ``varying only in a direction $e$''. In fact,  both functionals (Kac and Ginzburg-Landau)
are not convex, and the free energy of Kac model is not even local, so that direct methods [Da]
don't apply here.

Rearrangements methods were used in both [LiLo] for Ginzburg-Landau equation in a disc, and [AlBe]
for the scalar Kac model. Actually, only partial convexity is achieved when restricting to the
class of radially symmetric functions for Ginzburg-Landau, while for the scalar Kac model,
convexity is retrieved, by rearrangements, on a set of increasing functions. But it is hard to
figure out at least what {\it rearrangements} would mean in the vector (or complex) spin model. We
follow here an alternative route, elaborated in the papers [AlBeCasPr], [DeM], [DeMOrPrTr], and
culminating in [Pr], which consists in looking instead for extrema, verifying Euler-Lagrange
equation, as limiting orbits for a certain dynamics, known in that context as the {\it gradient
flow dynamics} $T_t$ (see (1.11)~). Free energy is a Lyapunov function for the gradient flow
dynamics, and thus we can resort on methods used in parabolic equations.

Our main results are related to existence of radially symmetric solutions $m_n(x)$ for
Euler-Lagrange equations $F_1(m)=0$ (see (1.8)~) associated to a suitable renormalized free energy
functional in infinite volume ${\cal F}_{\ren }$ defined in Proposition b.5. We have~:
\medskip
\noindent {\bf Theorem 0.1}: Assume that $J$ is on negative definite in the sense of [FrTo], and
the convolution operator expressed in the $\langle e_n\rangle$-sector enjoys asymptotic properties
described in Definition 1.1. Let $m(x)=(m_\beta+v(r))e^{in\theta}$, with $v\in \widetilde X^0_2$
(see (1.28)~) such that ${\cal F}_{\ren }(m)<\infty$. Then any limit point $m^*$ of $T_tm$,
$t\to\infty$ (in the sense of uniform convergence on the compact sets) satisfies $F_1(m^*)=0$ and
${\cal F}_{\ren }(m^*)\leq{\cal F}_{\ren }(m)$.
\smallskip
Thus we can describe some local minima in the space of all configurations $m$. Next we study the
spectrum of the $n$-th Fourier mode of the second variation operator $L_{m^*}$ around $m^*$. Here
we assume a particular form for $J$.
\medskip
\noindent {\bf Theorem 0.2}: Assume $J$ is a Gaussian, and moreover that the critical point $m^*$
verifies $m^*(x)=(m_\beta+v(r))e^{in\theta}$, with $v^*\in \widetilde X^0_2$. Then the $n$-th
Fourier mode of the second variation operator $L_{m^*}$ (due to breaking the gauge invariance) has
purely continuous spectrum.
\smallskip
(The decay property on $v^*$ could certainly be removed, see Sect.3.~) So we recover the result of
[OvSi] for the Ginzburg-Landau functional relative to the $n$-th mode only. Continuous,, but also
negative spectrum of $L_{m^*}$ near 0 suggests that only ``linear instability'' (not exponential)
may occur, in this mode, around the equilibrium. Nevertheless our conclusion remains much weaker
than for the Ginzburg-Landau functional [OvSi], since we have no information about higher modes.
Naive intuition suggests  that 0 is an eigenvalue  for the $n+1$-th Fourier mode, corresponding to
the breaking of the translation group. So the ``hedgehog conjecture'' described above in the case
of Ginzburg-Landau equation, remains largely open in case of the non local Kac functional (here the
case $n=1$ doesn't seem to play any special r\^ole).

\bigskip
\noindent{\bf 1. The gradient-flow dynamics}.
\medskip
We look for a solution of the free energy variational problem, among all configurations with given
degree $n\geq  1$, obtained as a limiting orbit of the gradient-flow dynamics. The minimization
problem in infinite volume involves an approximation process in finite boxes $\Lambda$, and
renormalization of the free energy. Definitions of the thermodynamic functions are given in
Appendix C, and we refer to [El-BoRo] for details.
\medskip
\noindent {\bf a) The infinite volume gradient flow dynamics}.
\smallskip
We start to construct the gradient flow dynamics in ${\bf R}^2$ subject to a vorticity condition at
infinity. It is  convenient to express everything in Fourier modes, using the identification
$L^2({\bf R}^2)\approx\oplus_{j\in{\bf Z}}e_j L^2({\bf R}^+;rdr)$, $e_j(\theta)=e^{ij\theta}$. In
Appendix B we discuss some properties of the degree, and show that higher harmonics can be involved
in the Fourier expansion of $m$, provided they decay sufficiently fast as $r\to\infty$~; but
because of non linearity, we choose a single component $j=n$, the $\langle e_n\rangle$ sector.

For given $n$, Euler-Lagrange equation for ${\cal F}_{\ren }$ given by (b.28) is the same as we
would obtain formally, i.e. without renormalizing the energy. It follows that it is of the form
$F_1(m)=0$, where
$$F_1(m)=-J*m+{1\over\beta}\widehat I'(|m|){m\over|m|}\leqno(1.8)$$
(recall the notation $\widehat I(|m|)=I(m)$ whenever $I$ is rotation invariant.~) Eqn.(1.8) is
equivalent to $F_0(m)=0$, where
$$F_0(m)=-m+f(\beta|J*m|){J*m\over|J*m|}\leqno(1.9)$$
Here $f=(\widehat I')^{-1}=I_1/I_0$, and $I_\nu$ denotes the modified Bessel function of order
$\nu$. Solutions of (1.8) or (1.9) tend to cluster near the manifold $|m|=m_\beta$, which is the
critical set for the free energy of mean field. Here $m_\beta>0$ for $\beta>2$ satisfies
$m_\beta=f(\beta m_\beta)$.

Let also ${\cal F}u(\xi)=\int_{{\bf R}^2}e^{-ix\xi}u(x)dx$ and $H_nu(\rho)=\int_0^\infty
J_n(r\rho)u(r)rdr$ denote Fourier and Hankel transformation, respectively. $H_n$ is defined on the
core $C_0^\infty({\bf R}^+)$ of smooth, compactly supported functions on the half line. For $u\in
C_0^\infty({\bf R}^+)$, we have $H_nu(\rho)={\cal O}(\rho^n)$ as $\rho\to 0$, and $H_nu(\rho)={\cal
O}(\rho^{-\infty})$ as $\rho\to\infty$. It extends [Ti] to a unitary operator on $L^2({\bf
R};rdr)$, and $H_n^*=H_n$.

The convolution operator thus becomes $J*=\bigl(H_n{\cal F}\widehat J H_n\bigr)_{n\in{\bf Z}}$. We
look for a radially symmetric solution of (1.9), of the form $m(x)=u(r)e^{in\theta}$, with $u(r)\to
m_\beta$ as $r\to\infty$, and $u(0)=0$. So we get, formally, $-A_nu+{1\over\beta}\widehat I'(u)=0$,
$A_n=H_n{\cal F}\widehat JH_n$, or
$$-u+f(\beta A_nu)=0\leqno(1.10)$$
Unless $\beta\leq2$, in which case $u=0$ is the unique solution of (1.10), and $m=0$ the unique
minimizer of ${\cal F}_{\ren }$ (see [El-BoRo,Proposition 3.5] for the case of a lattice,~) the map
$u\mapsto f(\beta A_nu)$ is not a contraction, so following [Pr], we are led to consider the
associated gradient-flow dynamics, and solve the ``heat equation''
$${du\over dt}=-u+f(\beta A_nu), \ u(0,r)=u_0\leqno(1.11)$$
Although $A_n$ is a bounded operator on $L^2({\bf R}^+;rdr)$, we are faced with the problem that
$u\notin L^2({\bf R}^+;rdr)$ because of the condition $u(r)\to m_\beta$ as $r\to\infty$. We need
the property, that $A_n$ acts naturally upon functions having asymptotics near infinity. It should
also map the constant function 1 to itself modulo $L^2$. So we introduce the~:
\medskip
\noindent{\bf Definition 1.1}: We say that $A_n$ has the asymptotic property iff, given $\chi$ a
smooth cut-off equal to 1 near infinity, for all $k\in{1\over2}{\bf Z}$, $A_n(\cdot)^{-k}\chi$ is a
smooth function on the half-line, with $A_n((\cdot)^{-k}\chi)(r)\sim r^{-k}(1+{a_1\over
r}+{a_2\over r^2}+\cdots)$, in the sense of asymptotic sums, as $r\to\infty$. Moreover, if $a_1=0$
we say that $A_n$ has the asymptotic property with vanishing subprincipal symbol.
\smallskip
We don't know if Definition 1.1 could actually be derived from similar arguments in Hankel's
theorem leading to the inversion of $H_n$ (see [Wa,p.458].~) Operator $A_n$ with the asymptotic
property reminds us of a Pseudo-Differential Operator, with $r$ as the fiber variable, and
``hidden'' phase coordinates that show up when writing the integral representation of Bessel
functions. It is non-local (not only because convolution is not local, but more seriously because a
change to polar coordinates always destroys locality,~) with a priori only weak decoupling from 0
to $\infty$, see Appendix A.

Since $f$ is only defined on ${\bf R}^+$, operator $A_n$ must be also positivity preserving, i.e.
$A_nu\geq 0$ almost everywhere (a.e.) if $u\geq 0$.
\smallskip
\noindent {\it Example 1}: $J$ is a gaussian normalized in $L^1$, so that ${\cal F}\widehat
J(\rho)=\exp[-p\rho^2]$, $p>0$~; we know [Wa,p.395]  that $A_n(r,r')$ is given by Weber second
exponential integral
$$\int_0^\infty \exp[-p\rho^2]J_n(r\rho)J_n(r'\rho)\rho d\rho={1\over 2p}\exp[-(r^2+r'^2)/4p]I_n(rr'/2p)\geq 0\leqno(1.12)$$
with equality only if $rr'=0$. So $A_n$ is clearly positivity preserving (more precisely,
positivity improving, i.e. $A_nu> 0$ a.e. if $u\geq 0$, not identically 0,~) and stationary phase
arguments, together with the asymptotic expansion  of modified Bessel function
$$I_n(x)={e^x\over\sqrt{2\pi x}}\bigl(1-(n^2-1/4)/(2x)+\cdots\bigr),\quad x\to\infty\leqno(1.13)$$
show that $r^kA_n((\cdot)^{-k}\chi)(r)\sim 1+(k^2-n^2)/r^2+\cdots$, so $A_n$ has the asymptotic
property with vanishing subprincipal symbol.
\smallskip
\noindent {\it Example 2}: ${\cal F}\widehat J(\rho)=e^{-p\rho}(1+p\rho)$, and $A_n(r,r')$ can be
computed taking derivatives with respect to $p$, from the expression [Wa,p.389]
$$\int_0^\infty {\exp[-p\rho]\over\rho}J_n(r\rho)J_n(r'\rho)\rho d\rho={1\over \pi\sqrt{rr'}}Q_{n-1/2}\bigl(
{r^2+r'^2+p^2\over 2rr'}\bigr)\leqno(1.14)$$ where $Q_{n-1/2}$ is a Legendre function of second
kind, for which we have the integral representation $Q_{n-1/2}(\cosh\eta)=\int_\eta^\infty(2\cosh
s-2\cosh \eta)^{-1/2}e^{-ns}ds$, see [Le,p.174]. It is easy to check again that $A_n(r,r')\geq 0$.
\smallskip
Note that these $J$'s are non negative definite in the sense of [FrTo], i.e. both $\widehat J$ and
${\cal F}\widehat J$ are $\geq 0$, this excludes antiferromagnetic potentials, or potentials with
non definite sign. Of course, ${\cal F}\widehat J\geq 0$ implies that $A_n$ is a positive operator
in the mean, i.e $(A_nv|v)\geq 0$ for all $v\in L^2({\bf R^+},rdr)$, but this is not sufficient for
our purposes. The additional condition $({\cal F}\widehat J)'(0)=0$ seems to ensure that $A_n$ has
the asymptotic property with vanishing subprincipal symbol.  These features are suggested again by
formal stationary phase expansions in the integral representation of Bessel functions, but from a
ill-behaved phase that becomes rapidly oscillating for large $r$.

Our first result deals with existence and regularity of solutions of (1.10) in $L^2$ or in Sobolev
spaces. This will not play the most important r\^ole in the sequel, but this is a very natural
property. We introduce the closed convex set $\widetilde W=\{v\in L^2({\bf R^+},rdr):
v(r)+m_\beta\geq 0  \ \ae \}$ and $\widetilde W([0,T])=C^0([0,T],\widetilde W)$ with norm
$$\|v(t,r);\widetilde W([0,T])\|=\sup _{t\in[0,T]}\|v(t,r)\|_{L^2}\leqno(1.18)$$
Note first that, since $\widehat J\geq 0$ and $\int J=1$, we have ${\cal F}\widehat J(\rho)\leq 1$
and $\|A_n\|\leq 1$, where $\|\cdot\|$ denotes the ($C^*$-) norm of operators.
\medskip
\noindent {\bf Theorem 1.2}: Let as above $A_n$ be positivity preserving, and enjoy the asymptotic
property of Definition 1.1 with vanishing subprincipal symbol. Then for all $T>0$ and all $v_0\in
\widetilde W$, there is a unique $v\in \widetilde W([0,T])$ such that $u=v+m_\beta$ verifies (1.11)
with $u|_{t=0}=v_0+m_\beta$.
\smallskip
\noindent {\it Proof}: Since $A_n$ is positivity preserving, $A_nu\geq 0$ a.e. if $u\geq 0$ a.e. so
(1.11) makes sense for $u=v+m_\beta$, $v\in \widetilde W$. Moreover, $A_nu=A_nv+v_1+m_\beta$,
$v_1=m_\beta(A_n1-1)\in L^2({\bf R}^+)$ because of the asymptotic property with vanishing
subprincipal symbol, and of Proposition a.1. Equation (1.11) with initial condition $v(0,r)=v_0(r)$
can be written in the integrated  form
$$v(t,r)=e^{-t}v_0(r)+\int_0^tdt_1e^{t_1-t}[f(\beta A_nv(t_1,r)+\beta v_1(r)+\beta m_\beta)-f(\beta m_\beta)]\leqno(1.19)$$
Let again $w(t,r)=v(t,r)-e^{-t}v_0(r)$, this rewrites as $w(t,r)=\Phi(t,r,w)$, with
$$\Phi(t,r,w)=\int_0^tdt_1e^{t_1-t}[f\bigl(\beta A_nw(t_1,r)+\beta m_\beta+\beta v_1(r)+\beta e^{-t_1}A_nv_0(r)\bigr)
-f(\beta m_\beta)]\leqno(1.20)$$ Given $v_0\in \widetilde W$, denote by $\widetilde W_{v_0}([0,T])$
the closed convex set of $C^0([0,T],L^2)$ consisting of all functions $w(t,r)$ such that
$w(t,r)+e^{-t}v_0(r)\in C^0([0,T],\widetilde W)$. Because $f\geq 0$, $\Phi$ maps $\widetilde
W_{v_0}([0,T])$ into itself. Moreover, since $\|f'\|_\infty=1/2$, the estimate
$$|\Phi(t,r,w_1)-\Phi(t,r,w_2)|\leq {\beta\over 2}\int_0^tdt_1e^{t_1-t}|A_n(w_1-w_2)(t_1,r)|\leqno(1.21)$$
and the fact that $\|A_n\|\leq 1$ on $\widetilde W$ prove that  if $\beta^2 T(1-e^{-2T})<8$, then
$\Phi$ is a contraction on $\widetilde W_{v_0}([0,T])$. Theorem 1.2 then follows from the group
property. $\clubsuit$.
\medskip
Consider next the derivatives of $u$. Applying the radial field $r\partial_r$ to
$J*m(x)=e^{in\theta}A_nu(r)$, $x=x_1+ix_2=re^{i\theta}$ we find, in the distributional sense~:
$$e^{in\theta}r{\partial\over\partial r}A_nu(r)=x_1{\partial J\over\partial x_1}*m(x)+
x_2{\partial J\over\partial x_2}*m(x)\leqno(1.22)$$ which shows, since $u=|m|$~:
$$|\partial_rA_nu(t,r)|\leq\|\nabla J\|_1\|u\|_\infty\leqno(1.23)$$
$\|\cdot\|_1$ being the $L^1$ norm. On the other hand, using (c.6) we compute
$$\bigl({\partial\over\partial r}+{n\over r}\bigr)A_n=B_n, \quad B_n(r,r')= \int_0^\infty d\rho\rho J_{n-1}(r\rho)J_n(r'\rho)
{\cal F}\widehat{\cal J}(\rho) \leqno(1.24)$$ where as before $B_n(r,r')$ stands for the kernel of
operator $B_n$ with measure $rdr$. We have
\medskip
\noindent {\bf Lemma 1.3}: With notations above, $\|B_n\|\leq 1$, $\|{1\over r}A_n\|\leq C$, for
some $C>0$.
\smallskip
\noindent {\it Proof}: The first inequality results from $B_n^*B_n=H_n({\cal F}\widehat{\cal
J})^2H_n$ and the fact that $H_n$ is unitary on $L^2({\bf R}^+,rdr)$. We use also (a.14) to write
$${1\over r}A_n={1\over 2n}(H_{n-1}\bigl(\rho{\cal F}\widehat{\cal J}(\rho)\bigr)H_n+
H_{n+1}\bigl(\rho{\cal F}\widehat{\cal J}(\rho)\bigr)H_n)$$ Since the multiplication by $\rho{\cal
F}\widehat{\cal J}(\rho)$ is bounded on $L^2$ we get the second inequality. $\clubsuit$
\medskip
From this we can extend Theorem 1.2 to show regularity of the $r$-derivative $u'$ of $u$. Namely,
let $\widetilde W^1$ be the closed convex set $\{v\in \widetilde W: v(r)+m_\beta\leq 1 \ \ae , \
v'\in L^2({\bf R}^+,rdr)\}$, and $\widetilde W^1([0,T])=C^0([0,T];\widetilde W^1)$ with Sobolev
norm as in (1.18). We have~:
\medskip
\noindent {\bf Proposition 1.4}: Let $v\in C^0({\bf R}^+,\widetilde W)$ be the solution of (1.11)
constructed in Theorem 1.2, with initial value $v_0\in\widetilde W^1$. Then for all $T>0$, $v\in
\widetilde W^1([0,T])$.
\smallskip
\noindent {\it Proof}: With notations as in the proof of Theorem 1.2, we have
$$\eqalign{
{\partial w\over\partial r}(t,r)&=\beta\int_0^tdse^{s-t}f'\bigl(\beta A_nw(s,r)+\beta m_\beta+\beta
v_1(r)+ \beta e^{-s}A_nv_0(r)\bigr)\cr &\times{\partial\over\partial
r}\bigl(A_nw(s,r)+v_1(r)+e^{-s}A_nv_0(r)\bigr)\cr }\leqno(1.25)$$ We denote by $\Phi'(t,r,w)$ the
RHS of (1.25) and prove that $\Phi'$ is a contraction on $\widetilde W^1([0,T])$ if $T>0$ is small
enough. But this results also from Lemma 1.3, and the fact that $w\in\widetilde W([0,T])$.
$\clubsuit$
\smallskip
This Proposition extends easily by induction to all derivatives, so Sobolev embedding theorem shows
that $r\mapsto v(t,r)$ inherits the regularity of its initial datum.

Existence and uniqueness result in $L^2$ however, falls far short of our needs to ensure existence
of a limiting orbit satisfying (a.10) as $t\to\infty$, or to provide suitable asymptotics,
essentially because we lack of a uniform bound on $\|v(t,r);\widetilde W([0,T])\|$ as $T\to\infty$.
We restrict henceforth to continuous functions that tend to 0 sufficiently fast as $r\to\infty$.
Such initial conditions will be used in the Barrier Lemma below. It turns out that the evolution
equation doesn't either provide a uniform bound on the $L^\infty$ norm of $u(t,r)$ but we can still
obtain indirectly such estimates. Asymptotic property of operators $A_n$ will be used as a hint to
model our functional spaces. So we consider the space $X^0_k$ of functions $v\in C_0({\bf R}^+)$,
such that $\sup_{r\in[1,+\infty[}|r^kv(r)|<\infty$. It is easy to see that $X^0_k$ is a separable
Banach space, with norm
$$\|v;X^0_k\|=\sup_{r\in[0,1]}|v(r)|+\sup_{r\in[1,+\infty[}|r^kv(r)|\leqno(1.28)$$
For $k=2$, we have $X^0_k\subset L^2$. In Appendix A, we give continuity properties of $A_n$ acting
on these spaces. Let also $\widetilde X^0_k$ be the closed convex subset of $X_k^0$ consisting of
functions $v\in C_0({\bf R}^+)$, such that $v(r)+m_\beta\geq 0$, and define $\widetilde
X^0_k([0,T])=C^0([0,T],\widetilde X_k^0)$.
\medskip
\noindent{\bf Theorem 1.5}: With the assumptions above, and if in addition $A_n$ has vanishing
principal symbol, then for all $T>0$ small enough, and all $v_0\in \widetilde X^0_k$, $k=1,2$,
there is a unique $v\in \widetilde X^0_k([0,T])$ such that $u=v+m_\beta$ verifies (1.11) with
$u|_{t=0}=v_0+m_\beta$.
\smallskip
\noindent{\it Proof}: Following the proof of Theorem 1.2, we need to check that $\Phi$ is a
contraction on $X^0_{k,v_0}([0,T])$, the closed convex set of $C^0([0,T];C_0({\bf R}^+))$
consisting of functions $w(t,r)$ such that $w(t,r)+e^{-t}v_0(r)\in C^0([0,T],\widetilde X^0_k)$.
First we estimate $|\Phi(t,r,w(r))|$, and write
$$|\Phi(t,r,w)|\leq{\beta\over 2}\int_0^tdt_1e^{t_1-t}\bigl(|A_nw(t_1,r)|+|v_1(r)|+e^{-t_1}|A_nv_0(r)|\bigr)\leqno (1.29)$$
By Proposition a.2, $A_n$ is a bounded operator on $X^0_k$, and since $v_1\in X^0_k$ by assumption,
we have $w\in\widetilde X_k^0$~; by the same remark as in the proof of Theorem 1.2, we conclude
that $\Phi(t,r,w)\in\widetilde X_{k,v_0}^0$. Then (1.23) shows that $\Phi$ is a contraction on
$\widetilde X^0_{k,v_0}([0,T])$, when $T>0$ is small enough, and the proof goes as in Theorem 1.2.
$\clubsuit$
\medskip
Again, by the group property, we find $v\in C^0({\bf R}^+,\widetilde X^0_k)$. From Theorem 1.5 we
can infer existence of limit points of the orbits~:
\medskip
\noindent{\bf Corollary 1.6}: With the same hypotheses as in Theorem 1.5, from any sequence
$t_n\to+\infty$, we can extract a subsequence $t_{n_j}$ such that $v(t_{n_j},r)\to v^*\in C^0({\bf
R}^+)$ as $j\to\infty$ for the convergence on compact sets.
\smallskip
\noindent{\it Proof}: The family $u(t,r)=m_\beta+v(t,r)$ is clearly bounded by 1, since
$m(t,x)=e^{in\theta}u(t,r)$, $x=re^{i\theta}$ solves the evolution equation corresponding to
(1.10). Consider next $w(t,r)$ as in the proof of Proposition 1.4, (1.23) shows that
${\partial\over\partial r}\bigl(A_n(w(s,r)+v_1(r)+e^{-s}A_nv_0(r)\bigr)$ is bounded uniformly in
$(t,r)$, so by integration of (1.25), $\partial_rw(t,r)$ is uniformly bounded, and so
$|\partial_rv(t,r)|\leq C$. It follows that the family $v(t,r)$ is also equicontinuous, and the
conclusion follows from Ascoli-Arzel\`a theorem. $\clubsuit$
\medskip
Let us extend once more our previous considerations. We enrich the structure of our Banach space
$X^0_k$ by requiring some asymptotic behavior near $\infty$. Consider indeed the set $Y^0_1$ of
functions $u\in X^0_1$, such that $ru(r)$ has a limit as $r\to\infty$, and if $\ell(u)=\lim
_{r\to\infty}ru(r)$, the function  $r(ru(r)-\ell(u))$ is bounded. It is easy to see that $Y^0_1$ is
a separable Banach space, with norm
$$\|u;Y^0_1\|=\sup_{r\in[0,1[}|u(r)|+\sup_{r\in[1,+\infty[}|ru(r)|+\sup_{r\in[1,+\infty[}
|r\bigl(ru(r)-\ell(u)\bigr)| \leqno(1.30)$$ Similarly, consider the set $Y^0_2$ of functions $v\in
X^0_2$, such that $rv(r)$ has a limit $\ell_0(v)$ as $r\to\infty$, $r^2v(r)-r\ell_0(v)$ has a limit
$\ell_1(v)$ as $r\to\infty$, and the function $r^{1/2}(r^2v(r)-r\ell_0(v)-\ell_1(v))$ is bounded.
It is easy to see that $Y^0_2$ is also a separable Banach space, with norm
$$\|v;Y^0_2\|=\sup_{r\in[0,1[}|v(r)|+\sup_{r\in[1,+\infty[}|r^2v(r)|+\sup_{r\in[1,+\infty[}
|r^{1/2}\bigl(r^2u(r)-r\ell_0(v)-\ell_1(v)\bigr)| \leqno(1.31)$$ By $\widetilde Y^0_k$, $k=1,2$ we
denote also as before the closed convex subspace of $Y_k^0$ consisting of functions $v$ such that
$v(r)+m_\beta\geq 0$, and $\widetilde Y^0_k([0,T])=C^0([0,T],\widetilde Y_k^0)$.
\medskip
\noindent{\bf Proposition 1.7}: With the assumptions of Theorem 1.5 then for all $T>0$ small
enough, and all $v_0\in \widetilde Y^0_k$, $k=1,2$, there is a unique $v\in \widetilde
Y^0_k([0,T])$ such that $u=v+m_\beta$ verifies (1.11) with $u|_{t=0}=v_0+m_\beta$.
\smallskip
The proof goes along the same steps as this of Theorem 1.5, but this time we need also Proposition
a.3 to control the last term in (1.30) or (1.31) for $\|v(t,\cdot);Y^0_k\|$. Note that we do not
expect asymptotics beyond this order, on account of our accuracy in estimating $A_n$.

Whatever the class $\widetilde X^0_k$ or $\widetilde Y^0_k$ to which the initial datum does belong,
we cannot ensure that $v^*$ itself belongs to this set, nor even to $C_0({\bf R}^+)$. Fortunately
the main properties we shall use don't hinge upon $v^*$ itself (except for Theorem 0.2, where we
are lead to make an hypothesis about the short range of $v^*$,~) but are rather inherited from
those of the finite time evolution $v(t,r)$, $t>0$. For this reason we shall also call a limiting
orbit of the flow a $X$-{\it ghost}, stressing that it proceeds from an initial datum in $X$.
Actually the phase picture for the whole dynamics may look quite complicated, unless we could prove
uniqueness of the limiting orbits of (1.11), at least for a given $u_0$.

Now we extend the gradient-flow dynamics to those positive functions, which are merely continuous
on ${\bf R}^+$ and bounded by 1. Indeed we shall eventually take a $\widetilde X^0_k$-ghost $u^*$
as a new initial datum, but use $u^*(t,r)$ only with $r$ in a compact set. We have~:
\medskip
\noindent{\bf Proposition 1.8}: Let $A_n$ as above be positivity preserving, but not necessarily
with the asymptotic property. Then for all $T>0$ small enough, and all $u_0\in C^0({\bf R}^+)$,
$\|u_0\|_\infty\leq 1$, there is a unique $u\in C^0({\bf R}^+)([0,T])$ such that $u(t,r)$ verifies
(1.11) with $u|_{t=0}=u_0$.
\smallskip
The proof is omitted, for it goes as in Theorem 1.5, without the additional requirements on
asymptotics at infinity. Iterates of $\Phi(t,r,w)$ to the fixed point converge uniformly for $r$ on
every compact of ${\bf R}^+$. Again, by the group property, we find $u\in C^0({\bf R}^+;C^0({\bf
R}^+))$.

Next we consider various comparison theorems. As in the scalar case [Pr], the comparison theorem or
maximum principle shows very useful in our situation. To fix the ideas, we state it for the full
dynamics. Recall that $u^+=v^++m_\beta, v^+\in C^0({\bf R}^+,\widetilde X^0_k)$ is  a supersolution
of the Cauchy problem (1.11) with initial datum $u^+_0\in \widetilde X^0_k$ if
$\|u^+(t,\cdot)\|_\infty\leq  1$, $u^+_0\geq u_0$  and verifies
$${du^+\over dt}\geq -u^+ + f(\beta A_nu^+)$$
We define analogously a subsolution $u^-$. Because $A_n$ is positivity improving, for all $t\geq
0$, $u^-(t,r)\leq u(t,r)\leq u^+(t,r)$. The first consequence of the maximum principle is that $u$
doesn't increase beyond $m_\beta$ if this holds for $u_0$. More precisely, we  can show  the
following~:
\medskip
\noindent {\bf Proposition 1.9}: Let $u(t,r)=v(t,r)+m_\beta$, $v\in C^0({\bf R}^+,\widetilde
X_k^0)$ be the solution of (1.11) with initial datum $u_0(r)$ satisfying $|u_0(r)|\leq \mu\leq 1$
for some $\mu\geq m_\beta$. Then $|u(t,r)|\leq \mu$ and for all $t\geq 0$.
\smallskip
The easiest way of proving this is to consider $m(t,r)=e^{in\theta}u(t,r)$, and the argument goes
as in as in [El-BRo, Proposition 3.3]. This holds equally for the partial dynamics. Another
consequence of the maximum principle is monotonicity~:
\medskip
\noindent {\bf Proposition 1.10}: Let $A_n$ be positivity preserving. If the initial datum $u_0(r)$
in (1.11) is an increasing function of $r$, then the same holds of $u(t,r)$ for all $t\geq 0$.
Assume moreover $A_n$ be positivity improving. If $u_0(r)$ is strictly increasing, then the same
holds of $u(t,r)$ for all $t\geq 0$.
\smallskip
\noindent {\it Proof}: Let $a>0$, and consider the new dynamics on $r\in]a,+\infty[$ given by
${du_a\over dt}=-u_a+f(\beta A_nu_a), \ u_a(0,r)=u_a$, where $u_a(r)=u_0(r-a)$ is the translate of
$u_0$. Since $u_a(r)\leq u_0(r)$, we see that $u$ is a subsolution of ${du\over dt}=-u+f(\beta
A_nu)$, so $u(t,r-a)\leq u(t,r)$ for all $t>0$. $\clubsuit$.
\smallskip
Note this implies the former result if we take $\lambda=m_\beta$ in Proposition 1.9. Of course
these properties extend, by continuity, to the limit points $u^*(r)$.

When considering the spectral problem in Sect.3, we shall need to know that the limiting orbits
belong also to $C^1({\bf R}^+)$. Taking the $r$-derivative of $u$ in (1.11) involves the derivative
of $A_nu$ which we can compute using (1.24), but it is very hard to give estimates on $B_n$ along
the lines of Appendix A, since $B_n$ doesn't enjoy the asymptotic property and so forth. Since we
will eventually take as an interaction the function $J$ of Example 1, we restrict to this case,
which is much simpler because the closed form of Weber second exponential integral involves just
one (modified) Bessel function $I_n$, instead of the correlations $J_n(r\rho)J_n(r'\rho)$.

As in (1.18), we consider the space $X^1_k$ of functions $v\in C^1({\bf R}^+)\cap X^0_k$, such that
$\sup_{r\in[1,+\infty[}|r^{k+1}v'(r)|<\infty$. The weight $r^{k+1}$ is chosen in such a way that we
can take the derivative of the asymptotics of $v$, and is consistent with the complete asymptotics
(1.13). It is easy to see that $X^1_k$ is a separable Banach space, with norm
$$\|v;X^1_k\|=\|v;X^0_k\|+\sup_{r\in[0,1]}|v'(r)|+\sup_{r\in[1,+\infty[}|r^{k+1}v(r)|\leqno(1.34)$$
For $k=2$, we have $X^1_k\subset H^1$ (the usual Sobolev space.~) Let also as before $\widetilde
X^1_k$ be the closed convex subset of $X_k^1$ consisting of functions $v\in C_0({\bf R}^+)$, such
that $v(r)+m_\beta\geq 0$, and define $\widetilde X^1_k([0,T])=C^0([0,T],\widetilde X_k^1)$. We
have~:
\medskip
\noindent {\bf Theorem 1.11}: With the assumptions above, for all $T>0$ small enough, and all
$v_0\in \widetilde X^1_k$, $k=1,2$, there is a unique $v\in \widetilde X^1_k([0,T])$ such that
$u=v+m_\beta$ verifies (1.11) with $u|_{t=0}=v_0+m_\beta$. Moreover, from any sequence
$t_n\to+\infty$, we can extract a subsequence $t_{n_j}$ such that $v(t_{n_k},r)\to v^*\in C^1({\bf
R}^+)$ as $j\to\infty$ for the convergence on compact sets.
\smallskip
\noindent {\it Sketch of the proof}: We argue as in Theorem 1.5, showing that for small $t>0$,
$\Phi(t,r,\cdot)$ is a contraction on $\widetilde X^1_{k,v_0}$. This follows from the fact that
$A_n$ (in the particular case where $J$ is a Gaussian) is a bounded operator on $\widetilde X^1_k$,
as can be shown by using the asymptotics at infinity of $I_n$ as in Proposition a.2~: for large
$rr'$ we replace using (1.13), $\exp[-(r^2+r'^2)/4\pi]I_n(rr'/2\pi)$ by
$(rr')^{-1/2}e^{-(r-r')^2/4\pi}$, and rely on standard gaussian integral arguments.

The last part of the Theorem follows as in Corollary 1.6 from the equicontinuity of the second
derivatives of $A_nu(t,r)$, as we can check by iterating (1.22), and Ascoli-Arzel\`a theorem.
$\clubsuit$.
\smallskip
Note that, since we proceed by extraction of subsequences, the $X^1_k$-ghost we obtain from the
sequence $v(t_{n_j},r)$ as a limiting orbit in the $C^1$-topology may not coincide with the
corresponding $X^0_k$-ghost in the $C^0$-topology obtained in Corollary 1.6, even with the same
initial datum. But the $X^1_k$-ghosts also enjoy the monotony property as in Proposition 1.10.

\medskip
\noindent{\bf b) The partial dynamics, and the Barrier Lemma}.
\smallskip
In order to cope with divergent integrals we need also study the dynamics in some finite boxes
$\Lambda=\{|x|\leq \lambda\}$, for which $\lambda\to\infty$. Outside $\Lambda$, the magnetization
$m$ is frozen to a configuration $m_{\Lambda^c}$ which acts as a boundary condition for the
evolution inside $\Lambda$. We follow again closely the main steps of [Pr]. Define the free energy
with boundary condition $m_{\Lambda^c}$ as
$${\cal F}(m_\Lambda|m_{\Lambda^c})={\cal F}_\Lambda(m_\Lambda)+{1\over 2}\int_\Lambda dx\int_{\Lambda^c}dx'J(x-x')
|m_\Lambda(x)-m_{\Lambda^c}(x')|^2\leqno(1.40)$$ where
$${\cal F}_\Lambda(m_\Lambda)={1\over 4}\int_\Lambda dx\int_\Lambda dx'J(x-x')
|m_\Lambda(x)-m_\Lambda(x')|^2+\int_\Lambda dx f_\beta(m_\Lambda(x))\leqno(1.41)$$ [Contrary to
Appendix C, we have removed from the second integral the term $f(\beta m_\beta)$, which amounts to
shift ${\cal F}_\Lambda(m_\Lambda)$ from a constant term, so long $\Lambda$ is kept fixed.~] Since
$\int J=1$, ${\cal F}(m_\Lambda|m_{\Lambda^c})<\infty$ for all $\lambda$. We take variations of
${\cal F}$ inside the sector $\langle e_n\rangle$, i.e. among all radially symmetric functions of
the form $m(x)=u(r)e^{in\theta}$. To simplify notations, remove subscript $n$ from $A_n$, and
define operators $A_\lambda$ and $A_\lambda^c$ by their kernel $A_\lambda(r,r')=\chi(r'\leq
\lambda)A(r,r')$ and $A_\lambda^c(r,r')=\chi(r'\geq \lambda)A(r,r')$ respectively (here $\chi$
denotes the (sharp) charateristic function.~) We denote also by $\widetilde u=u_\lambda\oplus
u_\lambda^c$ the function equal to $u_\lambda$ on $[0,\lambda]$ and $u_\lambda^c$ outside, and
write $A\widetilde u=(A_\lambda\oplus A_\lambda^c)(u_\lambda\oplus u_\lambda^c)=A_\lambda
u_\lambda+A_\lambda^c u_\lambda^c$. We shall use twidled $u$ and $v$ to refer to partial dynamics.
It follows from (1.40) that Euler-Lagrange equation for ${\cal F}(\cdot|m_{\Lambda^c})$ restricted
to the sector $\langle e_n\rangle$ is given by
$$F_1^\Lambda(\widetilde u)=-A\widetilde u+{1\over\beta}\widehat I'(u_\lambda)=0\leqno(1.42)$$
which is equivalent to
$$F_0^\Lambda(\widetilde u)=-u_\lambda+f(\beta A\widetilde u)=0\leqno(1.43)$$
So we define the partial dynamics by $T_t^\Lambda u_0=u_\lambda(t,\cdot)\oplus u_\lambda^c$ where
$u_\lambda(t,\cdot)$ solves
$${du_\lambda\over dt}=-u_\lambda+f(\beta A_\lambda u_\lambda+\beta A_\lambda^c u_\lambda^c), \quad u_\lambda(0,r)=u_0(r),
\quad 0\leq r\leq\lambda\leqno(1.44)$$ with the same initial condition $u_0(r)$ as for the full
dynamics (1.11). The integrated form of (1.44) is again
$$u_\lambda(t,r)=e^{-t}u_0(r)+\int_0^tdse^{s-t}f(\beta A_\lambda u_\lambda(s,r)+\beta A_\lambda^c u_\lambda^c(s,r)), \quad 0\leq r\leq
\lambda\leqno(1.45)$$ Existence and uniqueness for (1.45), expressed in terms of $\widetilde
v=v_\lambda\oplus v_\lambda^c$ for the decomposition $\widetilde u=(m_\beta +v_\lambda)\oplus
(m_\beta+v_\lambda^c)$ follows as before, and $v_\lambda\in C^0\bigl({\bf R}^+;\widetilde
W^1(\Lambda)\cap C^1(\Lambda)\bigr)$ [so long as we are concerned in the first $r$-derivative.~]
Compactness of the orbits results also from the uniform boundedness of $v_\lambda(t,r)$ and
$\partial_r v_\lambda(t,r)$. Namely, given any sequence $t_n\to\infty$, there is $\widetilde
u^*_\lambda\in C^1(\Lambda)$ such that $\lim_{k\to\infty} T_{t_{n_k}}^\Lambda \widetilde
u=\widetilde u^*$ for a subsequence $t_{n_k}\to\infty$.

Note that Propositions 1.9 and 1.10 also apply to partial dynamics (1.44). Decay properties of
$v_\lambda$ as $\Lambda\to\infty$ will follow from the ``Barrier Lemma'' (in the terminology of
[Pr],~) which is an essential tool in our analysis~; this compares the full dynamics $T_t$ as in
(1.11) with the partial dynamics $T_t^\Lambda$.
\medskip
\noindent {\bf Theorem 1.12}: Let $u(t,r)$ and $\widetilde u(t,r)$ solve respectively the full and
partial dynamics (1.11) and (1.44), with same initial condition $u_0\in \widetilde X^0_2$. Then for
all $T>0$, $u(t,r)-\widetilde u(t,r)\to 0$ and $\partial_ru(t,r)-\partial_r\widetilde u(t,r)\to 0$
uniformly for $0\leq r\leq\lambda$ and $0\leq t\leq T$, as $\lambda\to\infty$. [the $r$-derivative
of $\widetilde u(t,r)$ being understood almost everywhere, namely outside $r=\lambda$.]
\smallskip
\noindent {\it Proof}: We proceed somewhat as in the proof of [El-BoRo,Proposition 3.5]. Recall
that the restriction of $T_t^\Lambda$ to $[0,\lambda]$ is given by
$$u_\lambda(t,r)=e^{-t}u_0(r)+\int_0^tdt_1e^{-(t-t_1)}f(\beta A_\lambda u_\lambda+\beta A_\lambda^c u_\lambda^c)(t_1,r),\quad r<\lambda
\leqno(1.50)$$ where $u_\lambda^c$ is independent of $t$ and takes for instance the initial value
$u_0$. [Since we are not interested in uniqueness properties of the limiting orbits, we could
assume $u_\lambda^c=m_\beta$.~] Denote by $u(t,r)$ the solution of (1.11) on the full space and by
$\widetilde u(t,\cdot)=T_t^\Lambda u_0=u_\lambda(t,\cdot)\oplus u_\lambda^c$. So $\widetilde u$ is
piecewise continuous. We have
$$(u-\widetilde u)(t,r)=\chi(r<\lambda)\int_0^tdt_1e^{-(t-t_1)}g_\lambda(u,\widetilde u)(t_1,r)
+\chi(r>\lambda)(u-u_\lambda^c)(t,r)\leqno(1.51)$$ with $g_\lambda(u,\widetilde u)=f(\beta
Au)-f(\beta A_\lambda u_\lambda+\beta A_\lambda^c u_\lambda^c)$. Since $0<f'\leq 1/2$, we get
$$|g_\lambda(u,\widetilde u)(t,r)|\leq {\beta\over 2}
|A(u-u_\lambda^c)(t,r)|={\beta\over 2}
|A_\lambda(u-u_\lambda)(t,r)+A_\lambda^c(u-u_\lambda^c)(t,r)|\leqno(1.52)$$ all $r<\lambda$.
Proposition 1.3 shows that if $u_0-m_\beta\in\widetilde X_k^0$, then $u-m_\beta\in\widetilde
X_k^0([0,T])$ for all $T>0$, and the RHS of (1.52) is well defined. Applying $A=A_\lambda\oplus
A_\lambda^c$ to (1.51) we get in turn
$$A(u-\widetilde u)(t_1,r)=A_\lambda\int_0^{t_1}dt_2e^{t_2-t_1}g_\lambda(u,\widetilde u)(t_2,r)
+A_\lambda^c(u-u_\lambda^c)(t_1,r)$$ Using (1.52) and the fact that $A_\lambda$ is positivity
preserving, gives the estimate
$$|A(u-\widetilde u)|(t_1,r)\leq{\beta\over 2}\int_0^{t_1}dt_2e^{t_2-t_1}A_\lambda|A(u-\widetilde u)|(t_2,r)
+|A_\lambda^c(u-u_\lambda)|(t_1,r)$$ Inserting into (1.51) we  get
$$\eqalign{
&|u-\widetilde
u|(t,r)\leq\chi(r<\lambda)\bigl({\beta\over2}\bigr)^2\int_0^tdt_1e^{-(t-t_1)}\int_0^{t_1}dt_2e^{-(t_1-t_2)}
A_\lambda|A(u-\widetilde u)|(t_2,r)\cr &+\chi(r<\lambda){\beta\over2}\int_0^tdt_1e^{-(t-t_1)}
|A_\lambda^c(u-u_\lambda^c)|(t_1,r)+\chi(r>\lambda)|u-u_\lambda|(t,r)\cr }$$ Let
$T^{(k)}u(t)=\int_0^tdt_1\int_0^{t_1}dt_2\cdots \int_0^{t_{k-1}}dt_ku(t_k)$ denote the $k$-fold
integral of $u$, this formula can be carried over by induction as
$$\eqalign{
&|u-\widetilde u|(t,r)\leq\chi(r<\lambda)e^{-t}\bigl[\Sum_{j=1}^{k-1}\bigl({\beta\over2}\bigr)^j
T^{(j)}\bigl(e^{(\cdot)}A_\lambda^{j-1}|A_\lambda^c(u-u_\lambda^c)|\bigr)(t,r)\cr
&+\bigl({\beta\over2}\bigr)^k T^{(k)}\bigl(e^{(\cdot)}A_\lambda^{k-1}|A(u-\widetilde
u)|\bigr)(t,r)\bigr]+ \chi(r>\lambda)|u-u_\lambda^c|(t,r) }\leqno(1.54)$$ We first need an estimate
on $A_\lambda^c(u-u_\lambda^c)(t,r)$ for $r<\lambda$. We proceed as in Lemma c.1, using (c.6)
$$\eqalign{
&A_\lambda^c(u-u_\lambda^c)(t,r)=(n+1)\int_\lambda^\infty dr'(u-u_\lambda^c)(t,r')\int_0^\infty
d\rho {\cal F}\widehat J(\rho)J_n(r\rho)J_{n+1}(r'\rho)\cr &+\int_\lambda^\infty
dr'(u-u_\lambda^c)(t,r')\int_0^\infty d\rho\rho{\cal F}\widehat J(\rho)J_n(r\rho){d\over d\rho}
J_{n+1}(r'\rho)\cr }\leqno(1.55)$$ We split $\int_0^\infty d\rho {\cal F}\widehat
J(\rho)J_n(r\rho)J_{n+1}(r'\rho)$ into 2 parts, integrating respectively on $[0,1/\lambda]$ and
$[1/\lambda,\infty[$. For the first one, we make use of the bounds $|J_n(r\rho)|\leq C(r\rho)^n$,
$J_{n+1}(r'\rho)\leq C$, for the second part, of the bounds $|J_{n+1}(r'\rho)|\leq
C(r'\rho)^{-1/2}$, $J_n(r\rho)\leq C$, and of the rapid decrease of ${\cal F}\widehat J(\rho)$.
Altogether, we get with a new constant $C>0$~:
$$|\int_0^\infty d\rho
{\cal F}\widehat J(\rho)J_n(r\rho)J_{n+1}(r'\rho)|\leq C\bigl({1\over\lambda}\bigl(
\bigl({r\over\lambda}\bigr)^n\vee 1\bigr)+{1\over\sqrt {r'}}\bigr) \leqno(1.56)$$ Next we integrate
by parts $\int_0^\infty d\rho\rho{\cal F}\widehat J(\rho)J_n(r\rho){d\over d\rho} J_{n+1}(r'\rho)$,
the 2 first terms can be bounded as before, for the third term $\int_0^\infty d\rho\rho r{\cal
F}\widehat J(\rho)J'_n(r\rho)J_{n+1}(r'\rho)$ we split again according to
$[0,1/\lambda],[1/\lambda,\infty[$, for the first integral we use $r\rho J'_n(r\rho)|\leq
C\bigl((r\rho)^n\vee\sqrt{r\rho}\bigr)$, which gives the bound
$C{1\over\lambda}\bigl(\bigl({r\over\lambda}\bigr)^n\vee\sqrt{r\over\rho}\bigr)$, for the last 2
ones we use $|J'_n(r\rho)|\leq C$, which gives the bound $Cr/\sqrt {r'}$. Thus
$$|\int_0^\infty d\rho\rho{\cal F}\widehat J(\rho)J_n(r\rho){d\over d\rho} J_{n+1}(r'\rho)|\leq
C\bigl({1\over\lambda}\bigl(\bigl({r\over\lambda}\bigr)^n\vee\sqrt{r\over\rho}\bigr)+{r\over\sqrt
{r'}}\bigr)\leqno(1.57)$$ On the other hand, since the initial datum $u_0$ belongs to $\widetilde
X^0_2$, given $T>0$, by Theorem 1.5 there is $C_T>0$ (also depending continuously on $T$) such that
for all $t<T$, $|u-u_\lambda^c|(t,r)\leq C_T(1+r)^{-2}$. Inserting this estimate into (1.55), using
(1.56) and (1.57) we get by integration, for a new constant $C_T>0$
$$|A_\lambda^c(u-u_\lambda^c)(t,r)|\leq C_T{1\over\sqrt \lambda},\quad r\leq \lambda\leqno(1.58)$$
Next we notice that since $e^{in\theta}A_nu(r)=J*(e^{in\cdot}u)$, and $\int J=1$, we get
$|A_\lambda u|(r)\leq \sup_{[0,\lambda]}|u|$ for all $r>0$. It follows then from (1.58) that
$$A_\lambda^{j-1}|A_\lambda^c(u-u_\lambda^c)|(t,r)\leq C_T/\sqrt\lambda, \quad j=1,2,\cdots$$
Now we have again $|A(u-\widetilde u)(t,r)|\leq 1$ for all $t,r>$, and
$A_\lambda^{k-1}|A(u-\widetilde u)(t,r)|\leq 1$ for all $k$. Performing the successive integrations
we find that the series in (1.54) is uniformly convergent for $t$ in compact sets, so we can write,
for $0\leq r\leq \lambda$,
$$|u-\widetilde u|(t,r)\leq e^{-t}\Sum_{j=1}^{\infty}\bigl({\beta\over2}\bigr)^j
T^{(j)}\bigl(e^{(\cdot)}A_\lambda^{j-1}|A_\lambda^c(u-u_\lambda^c)|\bigr)(t,r)\leq C_T
e^{t(\beta/2-1)}{1\over\sqrt\lambda}
$$
which proves the first estimate on $u(t,r)-\widetilde u(t,r)$.

We cannot repeat this argument for the $r$-derivative, since $\partial_r$ doesn't commute with
$A_\lambda$ [while partial derivatives $\partial_{x_j}$ commute with convolution,~] so we proceed
indirectly. Take again $r$-derivative of (1.25), and iterate (1.22) once more to get
$|\partial_r^2A_nu(t,r)|\leq\|\nabla^2 J\|_1\|u\|_\infty$. By integration in the $s$ variable, we
find that $\partial^2_rw(t,r)$, and consequently $\partial^2_ru(t,r)$ are uniformly bounded for $t$
in compact sets. The same conclusion holds for the partial dynamics, showing that
$|\partial^2_r\bigl(u(t,r)-\widetilde u(t,r)\bigr)|\leq \Const $ uniformly for $t$ in compact sets.
Now we use the following well-known interpolation inequality~: for any bounded intervals
$K_1\subset\subset K_2$, there is $C>0$ such that for all $f\in C^2$ we have
$$\sup_{K_1}|f'|^2\leq C \sup_{K_2}|f|\bigl(\sup_{K_2}|f|+\sup_{K_2}|f''|\bigr)$$
Applying this to $f(r)=u(t,r)-\widetilde u(t,r)$ easily yields the required uniform estimate on
$\partial_r u(t,r)-\partial_r\widetilde u(t,r)$. The theorem is proved. $\clubsuit$

\bigskip
\noindent {\bf 2. Existence of a radially symmetric minimizer}.
\medskip
In this Section we study some continuity properties of the free energy in finite or infinite
volume~; continuity properties of the renormalized free energy are investigated in Appendix B.
Next, following [Pr], we identify the limit points $\widetilde u^*(r)$ in the box $\Lambda$ by
using the excess free energy functional ${\cal F}(m_\Lambda|m_{\Lambda^c})$, then we identify the
limit points $u^*(r)$ on the half-line, by using the renormalized energy ${\cal F}_{\ren }(m)$ in
the full space, and eventually prove Theorem 0.1.
\medskip
\noindent {\bf a) Continuity properties for the free energy}.
\smallskip
We start with general remarks on continuity properties for the free energy, which are quite close
to the case of configurations valued in $[-1,1]$ as in [Pr],  [AlBe],\dots. The only difference is
that we need to assume $|m|\neq 1$, since the entropy function $I(m)$ is unbounded as $|m|\to 1$.
Consider the functional
$${\cal F}(m)={1\over 4}\int_{{\bf R}^2}dx\int_{{\bf R}^2}dyJ(x-y)|m(x)-m(y)|^2+\int_{{\bf R}^2}dx f_\beta(|m(x)|)\leqno(2.1)$$
defined on the set $E=L^\infty({\bf R}^2;B_2(0,1))$ and valued in ${\bf R}^+\cup\{+\infty\}$.
\medskip
\noindent {\bf Proposition 2.1}: ${\cal F}$ is lower semicontinuous on $E$ with respect to
convergence almost everywhere.
\smallskip
\noindent {\it Proof}: Let $m\in E$, and $m_n\in E$ be a sequence converging a.e. to $m$.  Define
$$g_n(x)={1\over 4}\int_{{\bf R}^2}dyJ(x-y)|m_n(x)-m_n(y)|^2+f_\beta(|m_n(x)|)$$
and $g(x)$ similarly, with $m_n(x)$ replaced by $m(x)$. Expanding the square and using the
normalization of $J$ in $L^1$ we get
$$\eqalign{
&g_n(x)-g(x)={1\over 4}(|m(x)|^2-|m_n(x)|^2)+f_\beta(|m_n(x)|)-f_\beta(|m(x)|)\cr &-{1\over
2}\re(m_n(x)-m(x))J*\overline m(x)+{1\over 2}\re m_n(x)J*(\overline {m_n}-\overline m)(x)+{1\over
4} J*\bigl(|m_n|^2-|m|^2\bigr)(x)\cr}$$ The first term tends to 0 a.e., so does
$f_\beta(|m_n(x)|)-f_\beta(|m(x)|)$ since $f_\beta$ is continuous. Since $m\in L^\infty$ and $J\in
L^1$, $J*\overline m(x)$ is uniformly continuous and bounded, so again $(m_n(x)-m(x))J*\overline
m(x)$ tends to 0 a.e.. In the same way, by the dominated convergence theorem, $J*(\overline
{m_n}-\overline m)(x)$, tends to 0 locally uniformly in $x$, and this argument equally applies to
$J*\bigl(|m_n|^2-|m|^2\bigr)(x)$, proving that the last 2 terms in the decomposition of
$g_n(x)-g(x)$ above tend to 0 a.e.. Then Fatou lemma shows that
$$\lim\inf _{n\to\infty}{\cal F}(m_n)\geq{\cal F}(m)\leqno(2.2)$$
which proves the Proposition. $\clubsuit$
\medskip
Thus proving the existence of a minimizer for ${\cal F}$ amounts to extract from every minimizing
sequence a subsequence converging a.e.. The next result concerns weak lower semi-continuity of the
free energy in finite volume $\Lambda$ as is defined in (1.40) and (1.41). To simplify the
notations, we set $m=m_\Lambda$, and $m^c=m_{\Lambda^c}$. Here $m^c=m_{\Lambda^c}\in E$ is fixed.
\medskip
\noindent {\bf Proposition 2.2}: If $m_n\in E$ converges weakly to $m$ in $L^p(\Lambda)$, $1\leq
p<\infty$, then
$$\lim\inf _{n\to\infty}{\cal F}(m_n|m^c)\geq{\cal F}(m|m^c)\leqno(2.3)$$
while, if $|m_n(x)|\leq\mu<1$ and $m_n\to m$ a.e. in $\Lambda$, then ${\cal F}(m_n|m^c)\to{\cal
F}(m|m^c)$.
\smallskip
\noindent {\it Proof}: Following [Pr] we write ${\cal F}(m|m^c)=F(m|m^c)+R(m^c)$ where
$$\eqalign{
F(m&|m^c)={1\over\beta}\int_\Lambda dx I(m(x))-{1\over 2}\re\int_{\Lambda}dx \int_{\Lambda}dy
J(x-y)m(x)\overline{m(y)}\cr &-\re\int_{\Lambda}dx \int_{\Lambda^c}dy
J(x-y)m(x)\overline{m^c(y)}\cr &R(m^c)={1\over 2}\int_{\Lambda}dx \int_{\Lambda^c}dy
J(x-y)|m^c(y)|^2\cr }\leqno(2.4)$$ So $R(m^c)$ is just a constant. Because $m_n\in E$ converges
weakly to $m$ in $L^p(\Lambda)$, $\int_{\Lambda}dy J(x-y)\overline{m_n(y)}$ tends to
$\int_{\Lambda}dy J(x-y)\overline{m(y)}$ as $n\to\infty$ for any $x\in\Lambda$. Moreover ${\cal
I}_1(m)={1\over 2}\re\int_{\Lambda}dx \int_{\Lambda}dy$ $ J(x-y)m(x)\overline{m(y)}$ is weakly
continuous on $L^p(\Lambda)$~; namely, if $m_n$ converges weakly to $m$, then
$m_n(x)\overline{m_n(y)}$ converge to $m(x)\overline{m(y)}$ weakly* in
$L^\infty(\Lambda\times\Lambda)$, and since $J(x-y)$ belongs to $L^1(\Lambda\times\Lambda)$, then
${\cal I}_1(m_n)\to {\cal I}_1(m)$. We get the same conclusion for the term ${\cal
I}_2(m)=\int_{\Lambda}dx \int_{\Lambda^c}dy J(x-y)m_n(x)\overline{m^c(y)}$, so that (2.3) follows
from the convexity of the entropy function $I(m)=\widehat I(|m|)$.

If $m_n\to m$ a.e. in $\Lambda$, then by the dominated convergence theorem, the last 2 terms in
$F(m_n|m^c)$ tend to the corresponding ones with $m$ instead of $m_n$. The same conclusion holds
for the first term provided $|m_n(x)|\leq\mu<1$ in $\Lambda$. $\clubsuit$
\medskip
Following [AlBe,Remark 4.8] we can use the first part of Proposition 2.2 to prove existence of a
minimizer for ${\cal F}(\cdot |m^c)$ subject to some weakly closed constraint, since ${\cal
F}(\cdot |m^c)$ itself is not coercive. We can also extend Proposition 2.2 to infinite volume for
${\cal F}$ as in (2.1), to show that $E$ is weak* compact in $L^\infty({\bf R})^2$, and ${\cal F}$
is weak* lower semi-continuous on $E$. Thus we can prove again existence of a minimizer for ${\cal
F}$ subject to some weakly closed constraint~; but it turns out that a topological constraint such
as the degree of $m$ at infinity is not weakly closed, and thus we shall proceed another way.
\medskip
\noindent {\bf b) Free energy dissipation for the partial dynamics}.
\smallskip
This paragraph is a  first step towards Theorem 0.1. The key property of the solution of (1.11) or
(1.44) is that its energy decreases with time. To start with, we consider the case of finite
volume. Given $\Lambda$ we write again, emphasizing the dependence on $\theta$ in the sector
$\langle e_n\rangle$, the solution of the partial dynamics $T^\Lambda_t$ with initial value
$e^{in\theta}u_0(r)$  as $\widetilde m(t,x)=T^\Lambda_t(e^{in\theta}u_0(r))$, using the notation
$\widetilde m=m_\Lambda\oplus m_{\Lambda^c}=e^{in\theta}\widetilde u=e^{in\theta}(u_\lambda\oplus
u_\lambda^c)$. Sometimes we omit to write $e^{in\theta}$. We define the free energy dissipation
rate of $\widetilde u(t,r)$ as
$${\cal I}^\Lambda(\widetilde u)(t,r)={1\over\beta}\int_0^\lambda dr\ r\bigl(-\beta A\widetilde u
+\widehat I'(u_\lambda)\bigr)\bigl(u_\lambda-f(\beta A\widetilde u)\bigr)\leqno(2.8)$$ It is easy
to see that ${\cal F}(m_\Lambda|m_{\Lambda^c})$ is a Lyapunov function for Eqn. (1.44), i.e. ${\cal
I}^\Lambda(\widetilde u)\geq 0$ and
$${\cal F}\bigl((T_t^\Lambda\widetilde m)_\Lambda|m_{\Lambda^c}\bigr)-{\cal F}\bigl(m_\Lambda|m_{\Lambda^c}\bigr)=
-\int_0^tds{\cal I}^\Lambda(\widetilde u)(s,r)\leqno(2.9)$$ with ${\cal I}^\Lambda(\widetilde u)=0$
iff $\widetilde u$ verifies (1.44). If the initial datum $\widetilde u$ is bounded below from 1, so
is $u_\lambda$ because of Proposition 1.10, and $\widehat I'(u_\lambda)<\infty$ everywhere. This
shows that ${\cal I}^\Lambda(\widetilde u)(t,r)<\infty$. We have the following
\medskip
\noindent{\bf Theorem 2.3}: Let the initial datum $\widetilde m$ be such that ${\cal
F}(m_\Lambda|m_{\Lambda^c})<+\infty$. Then every limit point $\widetilde u^*(r)$ of
$T^\Lambda_t\widetilde u$ satisfies Euler-Lagrange Eqn. (1.42)-(1.43)  and
$${\cal F}(e^{in\theta}u^*_\lambda(r)|m_{\Lambda^c})\leq {\cal F}(m_\Lambda|m_{\Lambda^c})\leqno(2.10)$$

\smallskip
\noindent {\it Proof}: We follow an argument of [Pr,Sect.4.2.5], essentially due to [FiMc-L]. Since
$\widetilde u^*$ is a limit point of $T_t^\Lambda u_0$, there is a sequence $t_n\to\infty$ such
that $\lim _{n\to\infty}\|T_{t_n}^\Lambda u_0-\widetilde u^*\|_\infty=0$. If $\widetilde u^*$ does
not satisfy (1.42), then ${\cal I}^\Lambda(\widetilde u^*)>0$. We start from $\widetilde u^*$ as a
new initial datum for the evolution $T_t^\Lambda$. Because $t\mapsto T_t^\Lambda(\widetilde u^*)$
is continuous as a map ${\bf R}^+\to C^0(\Lambda)$, and ${\cal I}^\Lambda$ is continuous (hence
l.s.c) on $C^0(\Lambda)$, for $t>0$ sufficiently small we have~: ${\cal
I}^\Lambda(T_t^\Lambda\widetilde u^*)>0$ and hence the free energy dissipation in $[0,1]$ verifies
$$D^\Lambda(\widetilde u^*)=\int_0^1dt{\cal I}^\Lambda(T_t^\Lambda\widetilde u^*)>0\leqno(2.11)$$
We shall show in a while that  this implies
$$\lim _{t\to\infty}\int_0^tds{\cal I}^\Lambda(T_s^\Lambda u_0)=+\infty\leqno(2.12)$$
which contradicts hypothesis ${\cal F}(m_\Lambda|m_{\Lambda^c})<\infty$ by inequality $\leq$ in
(2.9)~:
$$\int_0^tds{\cal I}^\Lambda(T_s^\Lambda u_0)\leq
{\cal F}\bigl(m_\Lambda|m_{\Lambda^c}\bigr)-{\cal F}\bigl((T_t^\Lambda
e^{in\theta}u_0)_\lambda|m_{\Lambda^c}\bigr) \leq{\cal
F}\bigl(m_\Lambda|m_{\Lambda^c}\bigr)<\infty$$ So for $r\leq \lambda$ we have, $-\beta A\widetilde
u^* +\widehat I'(u_\lambda^*)=0$, or equivalently $u_\lambda^*-f(\beta A\widetilde u^*)=0$, and the
limiting orbit $\widetilde u^*$ verifies Euler-Lagrange equation (1.42). Then (2.10) easily follows
from the uniform convergence on compact sets, as $t\to\infty$, of $(T_t^\Lambda
e^{in\theta}u_0)_\lambda$ towards $e^{in\theta}u_\lambda^*$.

Now we need to show that (2.11) implies (2.12). Because of the continuous dependence of the initial
data, we have again $\lim_{n\to\infty}\sup_{t\leq 1}\|T_t^\Lambda\widetilde
u-T_t^\Lambda(T_{t_n}^\Lambda u_0)\|_\infty=0$. Since we work in the finite volume $\Lambda$, and
$T_t^\Lambda u_0$ is bounded below from 1, this implies $\lim_{n\to\infty}D^\Lambda(T_{t_n}^\Lambda
u_0)=D^\Lambda(\widetilde u^*)>0$. So there is $N\in{\bf N}$ such that $D^\Lambda(T_{t_n}^\Lambda
u_0)\geq\delta>0$ for $n\geq N$. Without loss of generality (since $t_n\to\infty$) we can assume,
after possibly extracting a subsequence, that $N=1$, $t_0=1$ and $t_j-t_{j-1}\geq 1$ for all $j\geq
1$, so $t_1\geq 2$ and by the group property we have
$$\int_1^{t_1}dt{\cal I}^\Lambda(T_t^\Lambda u_0)=\int_0^{t_1-1}dt{\cal I}^\Lambda(T_t^\Lambda T_1^\Lambda u_0)
\geq D^\Lambda(T_1^\Lambda u_0)\geq \delta$$ By induction we get when $t\to\infty$~:
$$\int_0^tds{\cal I}^\Lambda(T_s^\Lambda u_0)\geq \bigl(\int_1^{t_1}+\int_{t_1}^{t_2}+\cdots\bigr)
ds{\cal I}^\Lambda(T_s^\Lambda u_0)\geq\delta+\delta+\cdots\to\infty$$ which proves (2.12).
$\clubsuit$.
\medskip
\noindent {\bf c) Free energy dissipation in infinite volume, and proof of Theorem 0.1}.
\smallskip
We extend here the results of Paragraph b) to full dynamics. Let $m\in E$ be the initial condition
for the full dynamics, such that ${\cal F}_{\ren }(m)<\infty$. More precisely, we choose
$m(x)=e^{in\theta}u_0(r)$, $u_0=v_0+m_\beta\in{\cal W}$, $v_0\in\widetilde X^0_2$, with notations
of Theorems b.6 and 1.5. For $\lambda>0$, denote by $\widetilde m=m_\Lambda\oplus m_{\Lambda^c}$,
and $\widetilde u=u_\lambda+u_\lambda^c$ the corresponding radial parts. Here we shall let
$\lambda\to\infty$, for fixed $t$. Using that $(T_t^\Lambda \widetilde
m)_{\Lambda^c}=m_{\Lambda^c}$, Theorem b.6 and Remark b.7 easily show that we can rewrite (2.9) as
$${\cal F}_{\ren }\bigl(T_t^\Lambda\widetilde m)\bigr)-{\cal F}_{\ren }\bigl(m)\leq
-\int_0^tds{\cal I}^\Lambda(\widetilde u)(s,r)\leqno(2.15)$$ By the Barrier Lemma, Theorem 1.12, we
have $\|T_t^\Lambda\widetilde m-T_tm;L^\infty(\Lambda)\|=\sup_{r\in[0,\lambda]}|\widetilde
u(t,r)-u(t,r)|\to 0$, and $\|\partial_rT_t^\Lambda\widetilde
m-\partial_rT_tm;L^\infty(\Lambda)\|=\sup_{r\in[0,\lambda]}|
\partial_r\widetilde u(t,r)-\partial_ru(t,r)|\to 0$, as $\lambda\to\infty$ uniformly for $t\in[0,T]$.
By [lower-semi] continuity of ${\cal F}_{\ren }$, Theorem b.8
$$\lim\inf_{\lambda\to\infty}\bigl[{\cal F}_{\ren }(T_t^\Lambda\widetilde m)-{\cal F}_{\ren }(m)\bigr]\geq{\cal F}_{\ren }(T_tm)
-{\cal F}_{\ren }(m)\leqno(2.16)$$ On the other hand, Fatou Lemma shows that
$$\lim\inf_{\lambda\to\infty}\int_0^tds {\cal I}^\Lambda(\widetilde u)\geq\int_0^tds {\cal I}^\Lambda(T_s u_0)$$
where ${\cal I}(u)$ is defined as in (2.8) by integrating over $r\in[0,\infty[$ and takes its
values in $[0,+\infty]$. We rewrite this inequality as~:
$$\lim\sup_{\lambda\to\infty}-\int_0^tds {\cal I}^\Lambda(\widetilde u)\leq-\int_0^tds {\cal I}^\Lambda(T_s u_0)$$
so by (2.15)
$${\cal F}_{\ren }(T_tm)-{\cal F}_{\ren }(m)\leq-\int_0^tds {\cal I}^\Lambda(T_s u_0) \leqno(2.17)$$
Now let $m^*(x)=e^{in\theta}u^*(r)$ be a limit point of $T_tm$, $u^*\in C^0({\bf R}^+)$ as in
Corollary 1.6, namely suppose there is a sequence $t_n\to\infty$ such that for any $\lambda>0$,
$\lim_{n\to\infty}\sup_{r\leq\lambda}|T_{t_n}u_0(r)-u^*(r)|=0$. So this time we let $t\to\infty$,
for fixed $\lambda$. If $u^*$ does not satisfy (1.10), then there is $\lambda^*>0$ so that ${\cal
I}^{\Lambda^*}(u^*)>0$.

We start from $u^*$ as a new initial datum for the evolution $T_t$. Note that $u^*$ has a priori no
decay at infinity, but we can consider $T_tu^*$ on $r\in[0,\lambda^*]$ by Proposition 1.8, with
convergence on compact sets. Because $t\mapsto T_t(u^*)$ is continuous as a map ${\bf R}^+\to
C^0(\Lambda^*)$, and ${\cal I}^{\Lambda^*}$ is continuous (hence l.s.c) on $C^0(\Lambda^*)$, for
$t>0$ sufficiently small we have~: ${\cal I}^{\Lambda^*}(T_tu^*)>0$ and hence the free energy
dissipation for $(t,r)\in[0,1]\times[0,\lambda^*]$ verifies
$$D^*(u^*)=\int_0^1dt{\cal I}^{\Lambda^*}(T_tu^*)>0\leqno(2.19)$$
As in the proof of Proposition 2.3, we shall show that this implies
$$\lim_{t\to\infty}\int_0^tds{\cal I}^{\Lambda^*}(T_su_0)=+\infty\leqno(2.20)$$
On the other hand, inequality (2.17) yields $\int_0^tds{\cal
I}^{\Lambda^*}(T_su_0)\leq\int_0^tds{\cal I}(T_su_0)\leq{\cal F}_{\ren }(m)-{\cal F}_{\ren
}(T_tm)$. So by the discussion at the end of Appendix B, there is $C>0$ such that ${\cal F}_{\ren
}(T_tm)\geq -C$ for all $t>0$ large enough, and we see that (2.20) contradicts the hypothesis
${\cal F}_{\ren }(m)<\infty$.

Now we prove that (2.19) implies (2.20). Since $T_{t_n}u_0(r)\to u^*$ uniformly for
$(t,r)\in[0,1]\times[0,\lambda^*]$ as $n\to\infty$, and the flow $T_t$ depends continuously on the
initial data, we see that $T_t(T_{t_n}u_0)(r)-T_tu^*(r)\to 0$ as $n\to\infty$, uniformly for
$(t,r)\in[0,1]\times[0,\lambda^*]$, and by Fatou Lemma,
$$\lim_{n\to\infty}D^*(T_{t_n}u_0)\geq\int_0^1dt{\cal I}^{\Lambda^*}(T_tu^*)>0$$
So we may assume that $D^*(T_{t_n}u_0)\geq\delta>0$ for all $n$ large enough, and the proof goes
exactly as in Proposition 2.3. This shows (2.20) and brings the proof of Theorem 0.1 to an end.
$\clubsuit$

\bigskip
\noindent{\bf 3. Linear stability}.
\medskip
To start with, we recall from [OvSi1] some well known facts concerning symmetry breaking of Eqn.
(1.8) or (1.9). The symmetry group $G$ for these equations is given by $G={\bf R}^2\times
O^+(2)\times U(1)\times\Gamma$, where ${\bf R}^2$ acts as translations $m(x)\to m(x-h)$, $h\in{\bf
R}^2$, $O^+(2)$ as rotations $m(x)\to m(R^{-1}x))$, $R\in O^+(2)$, $U(1)$ as gauge transformation
$m(x)\to\lambda m(x)$, $\lambda\in U(1)\approx S^1$, and $\Gamma$ as ``conjugation of charge''
$m(x)\to\overline{m(x)}$.

By the symmetry group $G_m$ of a solution $m$, we mean the largest subgroup of $G$ which leaves $m$
fixed. Then the part of $G$ broken by $m$ is the coset $G/G_m$. If $H$ is a one-parameter sub-group
of $G$, we say that $H$ is  {\it preserved} (resp. {\it broken}) by $m$ if $h(m)=m$ for all $h\in
H$ (resp. $h(m)\neq m$ for all $h\in H, h\neq\Id$.~)

The subgroup of translations is never preserved by $m$, unless $m$ is a constant. The symmetry
group of a radially symmetric solution, i.e. $m(x)=e^{in\theta}u(r)$ is the discrete subgroup of
$O_{k/n}^+(2)\subset O^+(2)$ of rotations by the angles $2k\pi/n,k\in{\bf Z}$. Thus, $m$ breaks the
translation group, the rotation subgroup $O^+(2)/O_{k/n}^+(2)$, and the charge group. The symmetry
group for an equation of the form $F(m)=0$ allows to find elements in the kernel of its
linearization around some point $m$, i.e. solutions of $\langle dF(m),\xi\rangle=0$. Namely, let
$m$ be a solution of $F(m)=0$, breaking a one parameter subgroup $g(s)\in G$ (the symmetry group of
this equation). Let $\tau$ be the generator of $g(s)$. Then $\xi=\tau m$ solves the linearized
equation $\langle dF(m),\xi\rangle=0$.
\medskip
\noindent {\bf a) Linearisation of ${\cal F}_{\ren }$ around a radially symmetric solution}.
\smallskip
Next we examine the Hessian of ${\cal F}_{\ren }$ and look for relations between the gradients of
$F_1$ and $F_0$ defined in (1.8), (1.9). We take advantage of the fact that $F_j$, $j=0,1$, are
real analytic functions, real for real $m$, to write $F_j(m)$ instead of $F_j(m,\overline m)$, and
we will denote also $F_j(\overline m)=\overline{F_j(m,\overline m)}=F_j(\overline m,m)$, ${\partial
\overline F_1\over\partial m}(m)={\partial F_1\over\partial \overline m}(\overline m)$, etc \dots
Following [OvSi], we express the Hessian of ${\cal F}_{\ren}$ in these complex variables as
$$\Hess {\cal F}_{\ren}=\pmatrix{{\partial^2{\cal F}\over\partial m\partial\overline m}\kern 1pt
&{\partial^2{\cal F}\over\partial\overline m^2}\kern 1pt \cr {\partial^2{\cal F}\over\partial
m^2}\kern 1pt &{\partial^2{\cal F}\over\partial m\partial\overline m}\kern 1pt }=
\pmatrix{{\partial F_1\over\partial m}\kern 1pt&{\partial F_1\over\partial \overline m}\kern 1pt
\cr {\partial \overline F_1\over\partial m}\kern 1pt &{\partial \overline F_1\over\partial
\overline m}\kern 1pt} \leqno(3.1)$$ We have $\Hess{\cal F}_{\ren}(m)=\nabla_m F_1(m)$ (the
gradient in the real sense) and similarly, we define $\nabla F_0(m)$. Introduce the notations
$z=\beta J*m$, $\phi_0(z)=f(|z|){z\over\overline z}$, $\phi_1(m)=\widehat I'(|m|){m\over|m|}$, so
that $\phi_0$ and $\phi_1$ are inverse from each other.
\medskip
\noindent{\bf Lemma 3.1}: With the notations above
$$-\beta\nabla_z \phi_0(z)\nabla_m F_1(m)=-\Id +\beta\nabla_z \phi_0(z)J*\cdot=\nabla_m F_0(m)\leqno(3.2)$$
and in particular $\bigl(\beta\nabla_z \phi_0(z)\bigr)^{-1}=\nabla_m F_1(m)+J*$.
\smallskip
\noindent{\it Proof}: Again, we write $\phi_1(m)$ for $\phi_1(m,\overline m)$, $\phi_1(\overline
m)$ for $\phi_1(\overline m,m)$, and similarly for $\phi_0$. The ``upper-left'' matrix element of
$\nabla \phi_0(z)\nabla F_1(m)$ is given by
$$a_1=-{\partial \phi_0\over\partial z}(z)J*\cdot +{1\over\beta}\bigl[{\partial \phi_0\over\partial z}(z){\partial
\phi_1\over\partial m}(m)+ {\partial \phi_0\over\partial\overline z}(z){\partial
\phi_1\over\partial \overline m}(\overline m)\bigr]$$ On the other hand, differentiating the
identity $\phi_0\circ \phi_1=\Id $ we get
$$1= \partial_m (\phi_0\circ \phi_1)(m)={\partial \phi_0\over\partial z}(z){\partial \phi_1\over\partial m}(m)+
{\partial \phi_0\over\partial\overline z}(z){\partial \phi_1\over\partial \overline m}(\overline
m)$$ which leads to $\beta a_1=1-\beta{\partial \phi_0\over\partial z}(z)J*\cdot$. All other matrix
elements can be handled of this sort, using $\partial_{\overline m} (\phi_0\circ \phi_1)(m)=0$, and
two similar identities, obtained after permuting $z$ with $\overline z$, $m$ with $\overline m$. So
we proved
$$-\beta\nabla \phi_0(z)\nabla F_1(m)=-\Id +\beta\pmatrix{{\partial \phi_0\over\partial z}(z)\kern 1pt
&{\partial \phi_0\over\partial\overline z}(z)\kern 1pt \cr {\partial \phi_0\over\partial \overline
z}(\overline z)\kern 1pt &{\partial \phi_0\over\partial z}(\overline z)\kern 1pt
}J*\cdot\leqno(3.3)$$ and the Lemma easily follows. $\clubsuit$
\medskip
A direct computation also shows ${\partial \phi_0\over\partial z}(z)={1\over
2}\bigl({f(|z|)\over|z|}+f'(|z|)\bigr)$. Using recursion formulas between the derivatives of the
modified Bessel functions $I_0$ and $I_1$, we find for $f=I_1/I_0$~: $f'(t)=1-f^2(t)-{f(t)\over
t}$, so
$${\partial \phi_0\over\partial z}(z)={1\over 2}(1-f^2(|z|))\leqno(3.4)$$
Similarly
$${\partial \phi_0\over\partial \overline z}(z)={z\over2\overline z}\bigl(1-f^2(|z|)-2{f(|z|)\over|z|}\bigr)\leqno(3.6)$$
so that setting $b(|z|)=1-f^2(|z|)-2{f(|z|)\over|z|}$, we have
$$\nabla_z \phi_0(z)={1\over 2}\pmatrix{1-f^2(|z|)\kern 1pt
&{z\over\overline z}b(|z|)\kern 1pt \cr {\overline z\over z}b(|z|)\kern 1pt &1-f^2(|z|)\kern 1pt
}\leqno(3.7)$$ It follows that $\nabla_z \phi_0(z)$ is hermitean, and
$$\det \nabla \phi_0(z)=a(|z|)=f'(|z|){f(|z|)\over|z|}\leqno(3.8)$$
We notice that $1-f^2(|z|)>0$, and $b(|z|)=f'(|z|)-{f(|z|)\over|z|}\leq 0$ with equality only at
$z=0$, since $f(0)=0$ and $f$ is strictly concave on $[0,+\infty[$. Now by Lemma 3.1
$$\nabla F_0(m)=-\Id +{\beta\over 2}\pmatrix{1-f^2(|z|)\kern 1pt
&{z\over\overline z}b(|z|)\kern 1pt \cr {\overline z\over z}b(|z|)\kern 1pt &1-f^2(|z|)\kern 1pt
}J*\cdot\leqno(3.9)$$

Let now $m$ be a radially symmetric solution of $d{\cal F}_{\ren}=0$, as in Theorem 0. If
$m(x)=e^{in\theta}u(r)$ solves (1.8), then $f(|z|)=u(r)$, and we have
$b(|z|)=b(r)=1-u^2(r)-2{u(r)\over\widehat I'\circ u(r)}$, $a(r)=a(|z|)=f'\circ f^{-1}\circ
u(r){u(r)\over f^{-1}\circ u(r)}$ (here we have used the relation $|m|=f(|z|)$ to denote, somewhat
incorrectly, $a(|z|)$ by $a(r)$, $b(|z|)$ by $b(r)$.~) Then Lemma 3.1 shows again
$$\Hess{\cal F}_{\ren}(m)=\nabla_m F_1(m)=-{1\over 2\beta}B(r,\theta)+J*\cdot\leqno(3.10)$$
with
$$B(r,\theta)={1\over a(r)}\pmatrix{1-u^2(r)\kern 1pt
&-e^{2in\theta}b(r)\kern 1pt \cr -e^{-2in\theta}b(r)\kern 1pt &1-u^2(r)\kern 1pt }\leqno(3.11)$$
(cf. [OvSi1, formula (7.7)] for Ginzburg-Landau equation.~) As in [OvSi] we have the
\medskip
\noindent {\bf Lemma 3.2}: $\nabla F_1$ is symmetric for the scalar product
$\langle\xi,\eta\rangle=\re \int\overline\eta\xi$. In other words, $\re\int\overline\eta\nabla
F_1(\xi)=\re\int\overline{\nabla_1(\eta)}\xi$, where $\nabla F_1(\xi)$ is a shorthand for
$\langle\nabla F_1(x),(\xi,\overline\xi)\rangle$.
\medskip
Now we make a Fourier analysis of $\nabla F_1$. Taking polar coordinates as above, write
$\xi=\Sum_{k\in{\bf Z}}\xi_k(r)e^{ik\theta}$, $\xi_k\in{\bf C}$, and expand $J*\cdot$ in terms of
$A_k$. To account for the phase factors $e^{\pm2in\theta}$ in (3.11) we make a shift of indices and
introduce the mapping
$$\pi:(\xi,\overline\xi)=\bigl((\xi_k,\overline\xi_k)\bigr)_{k\in{\bf Z}}\mapsto\widehat\xi=
\bigl((\xi_k,\overline\xi_{2n-k})\bigr)_{k\in{\bf Z}}$$ which is unitary if the target space is
endowed with the inner product
$$\langle\widehat\xi,\widehat\eta\rangle=\re\langle\widehat\xi_n,\widehat\eta_n\rangle+\re\Sum_{k>n}
\langle{\xi_k\choose\overline\xi_{2n-k}},{\eta_k\choose\overline\eta_{2n-k}}\rangle \leqno(3.13)$$
We then define the real-linear operator $\widehat\nabla_m F_1$ on vector-valued functions
$\widehat\xi$ by $\widehat\nabla_m F_1=\pi\nabla_m F_1\pi^*$. As in [OvSi] we can easily show that
operator $\widehat\nabla_m F_1$ is block-diagonal of the form
$$\langle\widehat\nabla_m F_1,\widehat\xi\rangle=\bigl(\langle\widehat\nabla_m F_1^{(k)},{\xi_k\choose\overline\xi_{2n-k}}
\rangle\bigr)_{k\geq n}$$ where
$$\widehat\nabla_m F_1^{(k)}=\pmatrix{A_k-{1\over 2\beta a(r)}(1-u^2(r))\kern 1pt
&{b(r)\over 2\beta a(r)}\kern 1pt \cr {b(r)\over 2\beta a(r)}\kern 1pt &A_{2n-k}-{1\over 2\beta
a(r)}(1-u^2(r))\kern 1pt }=L_k\leqno(3.14)$$ So we have reduced the problem of linear stability
around the radially symmetric solution $m(x)=e^{in\theta}u(r)$ to the study of the spectrum of the
family of operators $(L_k)_{k\geq n}$. By the discussion at the beginning of this Section, we
already know some zero modes, corresponding to breaking of the rotation, gauge or translation
symmetry. However, these functions are not in $L^2$.
\medskip
\noindent{\bf b) Spectral properties of $L_n$}.
\smallskip
Conjugating with $R={1\over\sqrt2}\pmatrix{-1\kern 1pt&1\kern 1pt \cr 1\kern 1pt &1\kern 1pt }$, we
see that $L_n$ is unitary equivalent to
$$\widetilde L_n=\pmatrix{A_n-{\widehat I'(u)\over\beta u}\kern 1pt
&0\kern 1pt \cr 0\kern 1pt &A_n-{\widehat I''(u)\over\beta}\kern 1pt }\leqno(3.15)$$ Let
$V(r)={\widehat I'(u)\over\beta u}>0$, $W(r)={\widehat I''(u)\over\beta}>0$. These are smooth
functions on ${\bf R}^+$. Since $A_nu-{1\over\beta}\widehat I'(u)=0$, $U={u\choose 0}$ is a
solution of $\widetilde L_nU=0$ (the zero mode is due to breaking the gauge group) but of course
$u\notin L^2$. First we look at the spectrum of $A_n$.
\medskip
\noindent{\bf Lemma 3.3}: If ${\cal F}\widehat J\geq 0$ and $\widehat J\geq0$ (i.e. $J$ is non
negative definite in the sense of [FrTo],~) then
$]0,1[\subset\sigma_c(A_n)\subset\sigma(A_n)\subset [0,1]$. Moreover $\ker A_n=0$.
\smallskip
\noindent {\it Proof}: Recall $A_n=H_n{\cal F}\widehat JH_n$, so
$(A_n-\lambda)^{-1}=-{1\over\lambda}H_n \bigl(1-{{\cal F}\widehat J\over\lambda}\bigr)^{-1}H_n$. So
$(A_n-\lambda)^{-1}$ is bounded in operator norm for $\lambda<0$, if ${\cal F}\widehat J\geq 0$. On
the other hand, ${\cal F}\widehat J(\rho)\leq 1$ since $\widehat J\geq 0$ and $\int J=1$, so when
$\lambda>1$, $\bigl(1-{{\cal F}\widehat J(\rho)\over\lambda}\bigr)^{-1}$ is a Neuman series
converging in operator norm. It follows that $\sigma(A_n)\subset[0,1]$. Conversely, let
$\lambda\in]0,1[$. Since $\widehat J(0)=1$ and ${\cal F}\widehat J(\rho)\to 0$ as $\rho\to\infty$,
there is $\rho_\lambda>0$ such that ${\cal F}\widehat J(\rho_\lambda)=\lambda$, so take $v\in
L^2({\bf R}^+;rdr)$ such that $H_nv(\rho)=\chi(\rho_\lambda-\e <\rho<\rho_\lambda+\e )$, where $\e
>0$ is small enough. It is clear that $(A_n-\lambda)^{-1}v\notin L^2$, which shows that
$A_n-\lambda$ is not surjective.

At last, let $\psi\in L^2({\bf R}^+;rdr)$ such that $A_n\psi=0$. Then ${\cal F}\widehat
J(\rho)H_n\psi(\rho)=0$ a.e. and this implies in turn that $H_n\psi(\rho)=0$ a.e. for we cannot
have ${\cal F}\widehat J(\rho)=0$ on a set of positive measure since $\widehat J$ is smooth and
$\widehat J\geq 0$. So $\psi=0$ and $\ker A_n=0$. $\clubsuit$
\medskip
Consider now $A_n-V(r)$, and $A_n-W(r)$. We discuss first some properties of $V$ and $W$. Convexity
of the entropy function $\widehat I$ on $[0,1]$ and $\widehat I'(0)=0$ imply that $\widehat
I''(u)\geq 0$, and $V'(u)=\widehat I''(u)-{\widehat I'(u)\over u}\geq 0$. In the same way, it is
easy to see that $\widehat I'''(u)\geq 0$. Recall also that $\widehat I''(0)=2$, and $m_\beta$
satisfies the equation $\widehat I'(m_\beta)=\beta m_\beta$. Although this will not be needed, we
mention for completeness that $\widehat I''(m_\beta)=\bigl({I_0''\over I_0}(\beta
m_\beta)-m_\beta^2\bigr)^{-1}$.

On the other hand we learned from Sect.2 that $u$ increases from 0 to $m_\beta$ on ${\bf R}^+$.
Altogether, this shows that $V$ increases from $2/\beta$ to 1 on $r\in[0,+\infty[$ for $\beta>2$
(recall that there is nothing to prove when $\beta\leq 2$,~) and $W$ increases from $2/\beta$ to
$\widehat I''(m_\beta)$.

So in the sense of self-adjoint operators, we have $A_n-1\leq A_n-V(r)\leq A_n-2/\beta$, and Lemma
3.3 easily shows that $\sigma(A_n)\subset [-1,1-2/\beta]$. For convenience we shift the whole
spectrum by 1 by changing $V$ into $\widetilde V=V-1$, and consider now $A_n-\widetilde V$. For a
suitable interaction $J$ we shall prove that the spectrum of $A_n-V$ is purely continuous by using
a positive commutator. This is actually a simple variant of Mourre estimates. More precisely, we
assume that $\widehat J(r)=(2\pi)^2e^{-r^2/4\pi}$, or equivalently ${\cal F}\widehat
J(\rho)=e^{-\pi\rho^2}$ (see Example 1 of Sect.1.~)
\medskip
\noindent{\bf Lemma 3.4}: With $J$ as above we have $[J*\cdot,x\partial_x+\partial_xx]=-4\pi\Delta
J$, or in the $\langle e_n\rangle$-sector, $[A_n,r\partial_r+\partial_rr]=4\pi
H_n(\rho^2e^{-\pi\rho^2})H_n$.
\smallskip
\noindent {\it Proof}: We compute for a test function $\varphi$,
$[J*\cdot,x\partial_x+\partial_xx]\varphi(x)=-2\int dxJ(x-y)\langle x-y,\nabla \varphi(y)\rangle$.
When $J(x)=(2\pi)^2e^{-x^2/4\pi}$, $-2J(x-y)(x-y)=4\pi J'(x-y)$, so Green formula shows that
$[J*\cdot,x\partial_x+\partial_xx]=-4\pi\Delta J$. Taking Fourier transform and restricting to the
$\langle e_n\rangle$-sector gives the Lemma. $\clubsuit$
\medskip
Denote by $D=r\partial_r+\partial_rr$ the generator of dilations, and $C_n=4\pi
H_n(\rho^2e^{-\pi\rho^2})H_n$. So $C_n$ is a bounded, positive operator in the $L^2$ sense (and
also positivity improving as we can show by differentiating (1.12) with respect to $p$ at $p=\pi$~;
also $C_n$ cuts off both low and high frequencies. These properties however, will not be used in
the sequel~.)

On the other hand, $[-\widetilde V,D]=2r\widetilde V'(r)$, which is also a positive (and positivity
improving) operator, so we get~:
$$[A_n-\widetilde V,D]=C_n+2r\widetilde V'(r)\leqno(3.17)$$
From this and the energy identity
$$\bigl([A_n-\widetilde V,D]\psi|\psi\bigr)=(C_n\psi|\psi)+(2r\widetilde V'(r)\psi|\psi)\leqno(3.18)$$
it follows that $\psi\in{\cal D}(D)=\{\psi\in L^2({\bf R}^+;rdr): r\psi'(r)\in L^2\}$ cannot be an
eigenfunction of $A_n-\widetilde V$, for otherwise it would be supported at $r=0$ (and the range of
its frequencies would reduce to $\rho=0$.~) But since $D$ is not bounded on $L^2$ we shall first
perform a regularization. This is simpler than usual (with the laplacian instead of a convolution
operator) because $A_n$ itself is bounded. Namely as in [Mo] we use the following~:
\medskip
\noindent {\bf Lemma 3.5}: For real $\mu\neq 0$, let $R_\mu=(1+D/\mu)^{-1}$. Then for $|\mu|$ large
enough, $R_\mu$ is uniformly bounded on $L^2({\bf R}^+;rdr)$, and s-$\lim_{\mu\to\infty}R_\mu=\Id
$.
\smallskip
\noindent {\it Remark}: Actually, as in [Mo] we could strengthen the conclusion of the Lemma by
replacing $L^2({\bf R}^+;rdr)$ by ${\cal H}_k$, $k=0, \pm 1,\pm 2$. Here ${\cal H}_0=L^2$, ${\cal
H}_2$ denotes the domain of a second order differential operator (laplacian) on the half-line
$\Delta_\alpha=r^{-2}\bigl((r\partial_r)^2+r\partial_r-\alpha\bigr)$, $\alpha>0$, i.e. ${\cal
H}_2=\{\psi\in L^2: \Delta_\alpha\psi\in L^2\}$, ${\cal H}_1$ is the closure of $C_0^\infty({\bf
R}^+)$ for the Sobolev norm $\|\psi\|_{1,\alpha}=\bigl(\int_0^\infty drr
(|\psi|^2+|\partial_r\psi|^2+{\alpha\over r^2}|\psi|^2)\bigr)^{1/2}$, and are ${\cal H}_{-1},{\cal
H}_{-2}$ the dual spaces. Actually some care is needed because of the singularity at $r=0$. We
shall not use this however, but it could be also useful, if we want to drop the assumption about
the existence of an asymptotic of $u(r)$ as $r\to\infty$, to work also on weighted spaces of the
form $\langle r\rangle^kL^2({\bf R}^+;rdr)$.
\smallskip
Given Lemma 3.5, we can prove absence of eigenvalues for $A_n-\widetilde V$ as follows.
Substituting $D_\mu=DR_\mu$ for $D$ in the LHS of (3.18) we get~: $\bigl((A_n-\widetilde
V)D_\mu\psi|\psi\bigr)-\bigl(D_\mu(A_n-\widetilde V)\psi|\psi\bigr)=\bigl(D_\mu\psi|(A_n-\widetilde
V)\psi\bigr)=0$ since $(A_n-\widetilde V)$ is self-adjoint and $D_\mu\psi\in{\cal D}(A_n-\widetilde
V)=L^2$. Now by the definition of $R_\mu$, we have $[A_n-\widetilde V,D_\mu]=R_\mu[A_n-\widetilde
V,D]R_\mu$. Using that s-$\lim_{\mu\to\infty}R_\mu=\Id $ and $[A_n,D]$ is bounded gives
s-$\lim_{\mu\to\infty} R_\mu[A_n,D]R_\mu=[A_n,D]$. We are left with $R_\mu[A_n,\widetilde V]R_\mu$.
Taking $r$-derivative of (1.10) gives $u'(r)=\beta f'(\beta A_nu(r))(A_nu)'(r)$, and by the
explicit representation (1.12) of $A_n$
$$ru'(r)=(2\pi)^{-2}\beta f'(\beta A_nu(r))\bigl(r[A_n,r]u(r)-2\pi nA_nu(r)\bigr)\leqno(3.19)$$
The second term on the RHS is uniformly bounded for $r\in{\bf R}^+$, since $A_n$ is bounded on
$L^\infty$. We know by Theorem 1.11 that $u\in C^1$. According to hypothesis of Theorem 0.2, we
shall also assume that $u$ has an asymptotics of the form $u(r)=m_\beta+{c\over r}+\cdots$ as
$r\to\infty$. Since $A_n$ has the asymptotic property with vanishing subprincipal symbol,
$r[A_n,r]$ is bounded on $m_\beta+X^0_2$, so $r\mapsto r[A_n,r]u(r)$  is bounded on ${\bf R}^+$. It
follows from (3.19) that the multiplication by $r\widetilde V'(r)$ is bounded on $L^2$, and again
by Lemma 3.5, s-$\lim_{\mu\to\infty} R_\mu[A_n,\widetilde V]R_\mu=[A_n,\widetilde V]$. So we get
$\bigl([A_n-\widetilde V,D_\mu]\psi|\psi\bigr)\to \bigl([A_n-\widetilde V,D]\psi|\psi\bigr)$ as
$\mu\to\infty$, and looking at the RHS of (3.18) gives that $\psi=0$.

Absence of singular spectrum follows also from this regularization and Putnam-Kato theorem (see
[ReSi,Sect.XIII.7].~)

By the discussion before Lemma 3.4, the same argument applies to $A_n-W(r)$, so by (3.15), $L_n$
has purely continuous spectrum, which eventually achieves proving Theorem 0.2.
\medskip
\noindent {\bf c) Remarks on the linear stability of higher modes}.
\smallskip
Spectral analysis of operators $L_k$, $k\geq n+1$ is much harder, and we shall content to write the
formula  for $L_{n+1}$. For the scalar 1-d Kac model, it is known (see [Pr,Sect.(6.3.1)] that if
$L$ denotes the  corresponding linear operator around an instanton $m(x)$, i.e. a solution to
Euler-Lagrange equation standing for (1.8) or (1.9), then $L$ is bounded, its spectrum lies in
${\bf R}^-$, and $L$ has a 1-d kernel generated by $m'(x)$. Moreover there is a spectral gap
$\omega>0$, namely $(\psi|L\psi)\leq -\omega\|\psi\|^2$ for all $\psi$ such that $(m'|\psi)=0$.
This relies on the fundamental energy identity
$$(\psi|L\psi)=-{1\over 2}\int dxdyJ(x-y)m'(x)m'(y)\bigl({\psi(x)\over m'(x)}-{\psi(y)\over m'(y)}\bigr)^2$$
which of course doesn't hold in our case. Instead we write, after conjugating with the reflection
matrix $R$ as before, operator $L_{n+1}$ in the form~:
$$\widetilde L_{n+1}= \pmatrix{ A_{n+1}+A_{n-1}-V(r)\kern 1pt & A_{n-1}-A_{n+1}\kern 1pt \cr
A_{n-1}-A_{n+1}\kern 1pt & A_{n+1}+A_{n-1}-W(r)\kern 1pt }$$ The vector $U(r)={nu(r)/r\choose
u'(r)}$ corresponds to the translation mode, and solves $\widetilde L_{n+1}U(r)=0$. With the
Gaussian $J$, naive considerations on the matrix $[\widetilde L_{n+1},D]$ suggest it could enjoy
positivity properties, but still allowing for a non trivial kernel.

\bigskip
\centerline{\bf Appendix}

\noindent{\bf A) Continuity properties of $A_n$}.
\smallskip
We analyze in this section the action of $A_n$ on $X^0_k$ and $Y^0_k$.
\medskip
\noindent{\bf Lemma a.1}: Assume $J\in{\cal S}({\bf R}^2)$ (the Schwartz space of  rapidly
decreasing functions together with all derivatives). Let $\chi$ be a smooth cut-off equal to 0 on
$[0,1/2] $ and to 1 on $[1,\infty[$, then if $u$ is continuous on the half-line, we have
$$\leqalignno{
&\sup_{r\leq 1}|A_n(1-\chi) u(r)|\leq C\sup_{r'\leq 1}|u(r')|&(a.3)\cr &\sup_{r\geq
1}r^k|A_n(1-\chi) u(r)|\leq C\sup_{r'\leq 1}|u(r')|,\quad k=0,1,2.&(a.4)\cr &\sup_{r\leq 1}|A_n\chi
u(r)|\leq C\sup_{r'\geq 1/2}|u(r')|&(a.5)\cr }$$ Moreover, (a.4) holds true for $k=5/2$ and for
$k=3$ when $n\geq 2$.
\smallskip
\noindent{\it Proof}: For (a.3), we simply use the fact that Bessel functions (together with all
their derivatives,~) are bounded on ${\bf R}^+$ (for example $J_0(x)|\leq1$, and $|J_n(x)|\leq
1/\sqrt{2}$, all $n\geq 1$), and ${\cal F}\widehat J(\rho)$ rapidly decreasing. Next we check
(a.4). For $k=0$, the same argument apply. When $k=1$, from the resursion formula for Bessel
functions,
$$(n+1)J_{n+1}(x)=x\bigl(J_n(x)-J'_{n+1}(x)\bigr)\leqno(a.6)$$
we get
$$\eqalign{
&rA_n(1-\chi)u(r)=(n+1)\int_0^\infty dr'r'(1-\chi)u(r')\int_0^\infty d\rho
J_{n+1}(r\rho)J_n(r'\rho) {\cal F}\widehat J(\rho)+\cr &+\int_0^\infty
dr'r'(1-\chi)u(r')\int_0^\infty d\rho r\rho J'_{n+1}(r\rho)J_n(r'\rho){\cal F}\widehat J(\rho)\cr
}\leqno(a.7)$$ By the uniform bound on Bessel functions we just recalled, the first term is
estimated by the RHS of (a.4). For the second term, we integrate by parts with respect to $\rho$ in
the second integral, using that $J\in{\cal S}({\bf R}^2)$, which yields the same conclusion. So
(a.4) is proved when $k=1$. Consider at last the case $k=2$. Denote by $b_1(r,r')$ and $b_2(r,r')$
the two $\rho$-integrals in (a.7). We compute
$$\eqalign{
&rb_1(r,r')=(n+1)(n+2)\int_0^\infty {d\rho\over\rho} J_{n+2}(r\rho)J_n(r'\rho) {\cal F}\widehat
J(\rho)+\cr &+(n+1)\int_0^\infty d\rho J_n(r'\rho) {\cal F}\widehat J(\rho){d\over
d\rho}J_{n+2}(r\rho)\cr }\leqno(a.8)$$ To estimate the first term, we split the integration over
$[0,1]$ and $[1,\infty]$, on $[0,1]$ we use that $J_n(r'\rho)\leq\Const (r'\rho)^n$ since $r'$ is
bounded on supp $1-\chi$. Performing the integration, this brings a term ${\cal O}(r'^n)$. The
integral on $[1,\infty[$ can be directly evaluated. We estimate the  second term in $rb_1(r,r')$,
and also $rb_2(r,r')$ in the same way. So we proved (a.4) for $k=0,1,2$ and for $k=3$ the same
arguments apply when $n\geq2$.

Let us consider the case $k=5/2$. Call $b_{11}(r,r')$ and $b_{12}(r,r')$ the 2 terms in (a.8), and
compute
$$r^{1/2}b_{11}(r,r')=(n+1)(n+2)\int_0^\infty{d\rho\over\rho^{3/2}}(r\rho)^{1/2}J_{n+2}(r\rho)J_n(r'\rho){\cal F}\widehat J(\rho)
\leqno(a.8)$$ Since, for all $n$
$$J_n(x)\sim\sqrt2 ({\pi x})^{-1/2}\cos(x-\pi n/2-\pi/4),\ x\to\infty$$
$(r\rho)^{1/2}J_{n+2}(r\rho)={\cal O}(1)$ uniformly for $r\rho>0$. We split the integral in (a.8)
as $\int_0^1$ and $\int_1^\infty$. Using $J_n(r'\rho)\leq\Const (r'\rho)^n$ for $r'\in\supp \chi$,
$\rho\in[0,1]$, the $\int_0^1$-integral contributes for a constant times $r'^n\int_0^1d\rho
\rho^{n-3/2}={\cal O}(r'^n)$, while the $\int_0^\infty$-integral contributes for a constant. Hence
$r^{1/2}\int_0^\infty dr'r'(1-\chi)u(r')b_{11}(r,r')\leq C\sup_{r'\in[0,1]}|u(r')|$. The same
arguments show this is the case of the corresponding term with $r^{1/2}b_{12}(r,r')$, and also of
those arising from $rb_2(r,r')$. So (a.4) holds true when $k=5/2$.

\noindent We proceed analogously for (a.5), and compute $A_n((\cdot)^3v)(r)$, where
$v(r')=\chi(r')r'^{-3}u(r')$, so that $|v(r')|\leq \chi(r')r'^{-3}\sup_{r'\geq 1/2}|u(r')|$, and
$\chi(r')r'^{-3}$ is integrable.

We estimate first $a_1(r,r')=r'\int_0^\infty d\rho \rho J_n(r\rho)J_n(r'\rho){\cal F}\widehat
J(\rho)$ using (a.6), as $r'\rho J_n(r'\rho)=(n+1)J_{n+1}(r'\rho)+\rho{d\over
d\rho}J_{n+1}(r'\rho)$ which gives accordingly $a_1(r,r')=a_{11}(r,r')+a_{12}(r,r')$. In
$a_{12}(r,r')$ we integrate by parts as before, which gives bounded terms for $r\leq1$. Multiply
again by $r'$, and write $a_2(r,r')=r'a_1(r,r')=a_{21}(r,r')+a_{22}(r,r')$, with
$a_{21}(r,r')=(n+1)(n+2)\int_0^\infty{d\rho\over\rho} J_n(r\rho)J_{n+2}(r'\rho){\cal F}\widehat
J(\rho)$, and $a_{22}(r,r')$, a less singular term, containing also derivatives of
$J_{n+2}(r'\rho)$, which we integrate by parts. We multiply a last time by $r'$, use again (a.6)
and write $a_3(r,r')=r'a_2(r,r')=a_{31}(r,r')+a_{32}(r,r')$, with
$a_{31}(r,r')=(n+1)(n+2)(n+3)\int_0^\infty{d\rho\over\rho^2} J_n(r\rho)J_{n+3}(r'\rho){\cal
F}\widehat J(\rho)$, and $a_{32}$ a less singular term, containing also derivatives of
$J_{n+3}(r'\rho)$. We examine $a_{31}$ by splitting the integration over $[0,\infty[$ into
$[0,r'^{-1}]$, $[r'^{-1},1]$, and $[1,\infty]$. On $[0,r'^{-1}]$, we use the estimates
$J_n(r\rho)\leq\Const (r\rho)^n$, and also $J_{n+3}(r'\rho)\leq\Const (r'\rho)^{n+3}$ when $n=1$.
So the first integral resulting from that splitting is bounded. The integral over $[1,\infty[$ is
also clearly bounded. Consider at last the second integral, and the most difficult case, i.e.
$n=1$. This time, we estimate simply $J_{n+3}(r'\rho)$ by a constant, and write
$J_n(r\rho)\leq\Const (r\rho)$~; integrating with respect to $\rho$ gives a term ${\cal O}(r\log
r')={\cal O}(\log r')$, but we can still divide this by $r'^\alpha$, $0<\alpha<1$, since
$r'\chi(r')r'^{-3+\alpha}$ is also integrable. All other terms can be handled similarly. So the
Lemma is proved. $\clubsuit$
\medskip
Now we give some continuity properties of $A_n$ on the space $X^0_k$.
\medskip
\noindent{\bf Proposition a.2}: Assume $A_n$ has the asymptotic property and is positivity
preserving. Then $A_n$ is a bounded operator on $X^0_k$, $k=1,2$.
\smallskip
\noindent {\it Proof}: First we show that $r\mapsto A_nu(r)$ is continuous. Let $\chi$ be a smooth
cutoff as above, we write $A_nu(r)=A_n(1-\chi)u(r)+A_n\chi u(r)$. As $A_n(r,r')$ is smooth and
$(1-\chi)u$ of compact support, the first term is a continuous function of $r$. For the second
term, write as in the proof of Lemma a.1 when $k=2$,
$$A_n\chi u(r)=\int_0^\infty dr'r'{\chi(r')\over r'^3}\bigl(r'^2u(r')\bigr)
\int_0^\infty d\rho {\cal F}\widehat J(\rho)J_n(r\rho)r'\rho J_n(r'\rho)\leqno(a.10)$$ and use(a.6)
to show that the $\rho$-integral is a continuous function of $(r,r')$, uniformly bounded with
respect to $r'\in[0,\infty[$. Since $u\in X_2^0$, $r'^2u(r')$ is also bounded on $\supp\chi$, and
$r'{\chi(r')\over r'^3}$ integrable, we can conclude that $A_n u$ is continuous.

When $k=1$, replace (a.10) by
$$A_n\chi u(r)=\int_0^\infty dr'r'{\chi(r')\over r'^3}\bigl(r'u(r')\bigr)r'
\int_0^\infty d\rho {\cal F}\widehat J(\rho)J_n(r\rho)r'\rho J_n(r'\rho)$$ and apply the procedure
above. We shall have to estimate $\int_0^\infty{d\rho\over\rho}J_n(r\rho)J_{n+2}(r'\rho)\widehat
J(\rho)$, for $r$ in a compact set we split again the integration over $[0,\infty[$into
$[0,\rho_0]$ and $[\rho_0,\infty]$, using the bound $J_n(r\rho)\leq\Const (r\rho)^n$ for the first
part, and absolute convergence for the second part. Thus we can conclude to continuity as in the
case $k=2$.

Next we show that if $u\in X^0_k$, then $r^kA_nu$ is bounded near infinity. Write
$r^kA_nu(r)=r^kA_n(1-\chi)u(r)+r^kA_n\chi u(r)$, by (a.4) we estimate the first term by
Const.$\|u;X^0_k\|$. For the second term, we have for all $r\geq1$,$|u(r)|\leq\|u;X^0_k\|r^{-k}$,
so using the fact that $A_n$ has the asymptotic property and is positivity preserving, we get
$|A_nu(r)|\leq\|u;X^0_k\|A_n(\cdot)^{-k}(r)\leq\Const r^{-k}\|u;X^0_k\|$. This brings the proof to
an end. $\clubsuit$

Along the same lines, we can prove continuity of $A_n$ in the asymptotic space $Y_k^0$.
\medskip
\noindent{\bf Proposition a.3}: Assume $A_n$ has the asymptotic property and is positivity
preserving. Then $A_n$ is a bounded operator on $Y^0_k$, $k=1,2$.

Moreover, if $A_n$ has vanishing subprincipal symbol, then it preserves the closed subspace
$E_0=\{u\in X^0_k : \ell_0(u)=0\}$.
\smallskip
\noindent {\it Proof}: Consider first $k=1$. We just need to show that $rA_nu(r)$ has a limit as
$r\to\infty$ (the same as $u$,~) and that
$$\sup_{r\in[1,+\infty[}|r\bigl(rA_nu(r)-\ell(u)\bigr)|\leq C\|u;Y^0_1\|\leqno(a.12)$$
For the first point, we observe that by definition, given $\e >0$, $-{\e \over r'}\leq
u(r')-{\ell(u)\over r'}\leq{\e \over r'}$ for $r'$ large enough. Multiply this inequality by
$\chi(r')$ as before, use that $A_n$ is positively preserving, and $A_n\bigl({\chi\over
r'}\bigr)(r)\sim {1\over r}$, we find $|A_n\chi u(r)-{\ell(u)\over r}|\leq{2\e \over r}$ for $r$
large enough. On the other hand, by (a.4) for $k=2$, we have $A_n(1-\chi)u(r)={\cal O}(r^{-2})$,
$r\to\infty$, so we can conclude $\lim_{r\to\infty}rA_nu(r)=\ell(u)$. It is also clear that if
$A_n$ has vanishing subprincipal symbol, then it preserves the closed subspace $E_0=\{u\in X^0_k :
\ell_0(u)=0\}$. Now we prove (a.12). The same argument as before shows that if $u\in Y^0_1$, then
$|A_n\chi u(r)-\ell(u)A_n{\chi\over r'}(r)|\leq C\|u;Y^0_1\|A_n{\chi\over r'^2}(r)$, so by the
asymptotic property, for $r$ large enough, $r^2|A_n\chi u(r)-{\ell(u)\over r}|\leq
C(\|u;Y^0_1\|+|\ell(u)|)\leq C'\|u;Y^0_1\|$. Also by (a.4), we have for $r\geq1$, $r^2|A_n(1-\chi)
u(r)|\leq C\sup_{r'\leq1}|u(r')|\leq C\|u;Y^0_1\|$. This proves (a.12) and the continuity property
for $k=1$.

The case $k=2$ goes similarly, taking advantage that (a.4) holds for $k=5/2$. $\clubsuit$
\smallskip
Our next result concerns the behavior of $A_nu(r)$ as $r\to0$.
\medskip
\noindent{\bf Proposition a.4}: If if $u\in X^0_k$, then $r^{-n}A_n(r)$ is bounded near 0..
\smallskip
\noindent{\it Proof}: We use this time the relation
$${2n\over x}J_n(x)=J_{n-1}(x)+J_{n+1}(x)\leqno(a.14)$$
which gives, by induction $2^{2p}(2p)x^{-2p}J_{2p}(x)=\Sum_{j=-p}^p\alpha_jJ_{2(p+j)}(x)$, $n=2p$,
and $2^{2p+1}(2p+1)x^{-2p-1}J_{2p+1}(x)=\Sum_{j=-p}^{p+1}\beta_jJ_{2(p+j)}(x)$, $n=2p+1$. where
$\alpha_j$ and $\beta_j$ are rational numbers. Then we argue as in Proposition a.2, for a linear
combination of terms of the form
$$\int_0^\infty dr'r'u(r')\int_0^\infty d\rho\rho^{n+1}J_{2p+2j}(r\rho)J_n(r'\rho){\cal F}\widehat J(\rho)$$
(we have assumed here $n=2p+1$ the case $n=2p$ is similar,~)and split accoording to the partition
of unity $\chi$ and $1-\chi$. The latter part is obviously bounded uniformly as $r\to0$. For the
$\chi$-integrals, we divide by $r'^3$ as in the proof of Proposition a.2. This easily gives the
Proposition. $\clubsuit$
\medskip
A more difficult problem would be to analyse the asymptotic behavior of $A_nu(r)$, as $r\to0$.
Formal analogy with radially symmetric solutions of the Ginzburg-Landau equations would suggest
that if $u(r)\sim r^n$ as $r\to0$, then so does $A_nu(r)$.

\bigskip
\noindent{\bf B. Renormalized free energy in infinite volume}.
\smallskip
In this section we renormalize (c.31) in infinite volume on the real plane, restricting $m$ to a
class of functions having topological degree at infinity. As usual, the divergence is due to the
self-energy corresponding to a neighborhood of the diagonal in $\int dx dyJ(x-y)|m(x)-m(y)|^2$, and
to $\int dx f_\beta(m(x))$. For the latter, we shall remove $f_\beta(m_\beta)$ from $f_\beta(m(x))$
provided $m$ stays sufficiently close to $m_\beta$. Our renormalized energy needs also to be
bounded from below.

We discuss first some properties of the degree of a complex valued function. See [FoGa] for more
advanced results. Let $m:{\bf R}^2\to {\bf C}$ be a differentiable function, considered as a vector
field on ${\bf R}^2$,  and subject to the condition $|m(x)|\to m_\beta$ as $|x|\to\infty$ uniformly
in $\widehat x=x/|x|$. Then the integer
$$\deg _{R} m={1\over 2\pi}\int_{|x|=R}d(\arg m)={1\over 2\pi}\int_{|x|=R}{dm\over m}\leqno (b.1)$$
independent of $R$ when $R>0$ is large enough, is called the (topological) degree of $m$ at
infinity, and denoted by $\deg _\infty m$. We do not attempt to characterize all functions
satisfying (b.1), but just state for completeness the following result (the condition on the $v_k$
could certainly be weakened. ~) We say that $m$ have a one-sided Fourier series if there is
$n\geq0$ (the case $n\leq 0$ follows by complex conjugation,~) such that
$m(x)=v_n(r)e^{in\theta}+\Sum_{k\geq n+1}v_k(r)e^{ik\theta}$.
\medskip
\noindent{\bf Proposition b.1}: Let $m$ have a one-sided Fourier series, $v_n(r)\to m_\beta$ and
$m(x)-v_n(r)e^{in\theta}\to 0$ as $r\to\infty$ uniformly in $\theta$. Assume moreover the Fourier
coefficients decay sufficiently fast to ensure existence of a continuous limit of $\Sum_{k\geq
n+1}v_k(r)z^k$ as $|z|\to 1^-$, i.e. $\Sum_{k\geq n+1}k^2|v_k(r)|^2<\infty$, uniformly as
$r\to\infty$. Then $\deg _\infty m=n$.
\smallskip
\noindent{\it Proof}: By assumption, we can differentiate the series term by term on the circle of
center 0, and radius $R$. To compute ${1\over 2\pi}\int_{|x|=R}{dm\over m}$, we put
$z=e^{i\theta}$. Since $m(x)-v_n(r)e^{in\theta}\to 0$ as $r\to\infty$ uniformly in $\theta$,
$|\Sum_{k\geq n+1}{v_k(R)\over v_n(R)}z^{k-n}|<1$ on $|z|=1$ for $R$ large enough, and Rouch\'e
theorem asserts that $v_n(R)+\Sum_{k\geq n+1}v_k(R)z^{k-n}$ has no zeroes inside the unit disc. So
the only pole inside the unit disc is $z=0$ and the corresponding residue is equal to $n$.
$\clubsuit$
\medskip
Unless restricting to those $m$ with $\deg _\infty m=0$, ${\cal F}(m|m^c)$ doesn't  make sense in
the limit $\Lambda\to\infty$, so we need a renormalization to remove the logarithmic singularity.
More precisely we show how to renormalize ${\cal F}(m)$, with ${\cal F}(m)$ as in (2.1) and  $m$ of
degree $n$. For simplicity we restrict to the case where $m$ has a single Fourier mode, i.e. $m$
belongs to the sector $\langle e_n(\theta)\rangle$. The idea is to remove in (2.1) a thin conic
neighborhood of the diagonal $x=y$ in ${\bf R}^2$ from the double integral.

To give some flavor of the general argument, we forget about the free energy density of the mean
field $f_\beta(|m|)$, and start with the particular case where $m(x)=m_\beta e^{in\theta}$, which
belongs to $E$ defined in (2.1) but of course, is not continuous at 0. Denote by  ${\cal
F}^0(m)={1\over 4}\int dx dyJ(x-y)|m(x)-m(y)|^2$ the interaction term. Let first $r_0>0$ and split
$$\eqalign{
&{\cal F}^0(m)={1\over 4}\bigl(\int_{|x|>r_0}dx\int_{|y|>r_0} dy\cr
&+\int_{|x|>r_0}dx\int_{|y|<r_0} dy+\int_{|x|<r_0}dx\int_{{\bf R}^2} dy\bigr)J(x-y)|m(x)-m(y)|^2\cr
}\leqno(b.9)$$ We are going to reduce (b.9) modulo integrable terms. Because $J$ decays rapidly at
infinity, it is easy to see that the last 2 terms are bounded. Consider then the first term, and
for each $x$ in the domain of integration, let $L_x=\{y\in {\bf R}^2: |y|>r_0, |ty+(1-t)x|>r_0,
\forall t\in[0,1]\}$ denote the {\it light cone} issued from $x$, and $S_x=\{|y|>r_0\}\setminus
L_x$ the corresponding {\it shadow cone}. Using the rapid decrease of $J$, we see that
$\int_{|x|>2r_0}dx\int_{S_x} dy|m(x)-m(y)|^2$ is bounded. We compute ${1\over 4}\int_{|x|>r_0}
dx\int_{|y|>r_0} dyJ(x-y)|m(x)-m(y)|^2$ by choosing polar coordinates $x=re^{i\theta}$, $x-y=\rho
e^{i\varphi}$, the volume element is $|dx\wedge d\overline x\wedge  dy\wedge  d\overline y|=4\rho
r|dr\wedge  d\theta\wedge  d\rho\wedge d\varphi|$, and now we perform the integration over
$\Omega=\{\rho\geq 0, r\geq r_0, \varphi\in[0,2\pi], \theta\in[0,2\pi]\}$. Writing
$y=r'e^{i\theta'}$ we have the relation
$$r'\sin(\theta'-\varphi)=r\sin(\theta-\varphi)\leqno(b.10)$$
Thus the first term in (b.9) rewrites as
$${1\over 4}\int_{|x|>r_0} dx\int_{|x|>r_0} dyJ(x-y)|m(x)-m(y)|^2=2m_\beta^2\int_\Omega  d\varphi  \rho d\rho \widehat J(\rho)
r dr  d\theta\bigl(1-\cos n(\theta-\theta')\bigr)\leqno(b.11)$$ We make the following
observations~: for fixed $\varphi$, let $\Gamma_\varphi$ be the reflection on the line
$\theta=\varphi$. Then for all $r>\rho$, the map $(r,\theta)\mapsto (r',\theta')$, is
$2\pi$-periodic in $\theta$ (it corresponds to shifting the circle of center 0 and radius $r$ by
the vector $\rho(\cos\varphi,-\sin\varphi)$.~) For given $r$, the function $\theta\mapsto r'$ is
even under $\Gamma_\varphi$, and increases from $r-\rho$ for $\theta=\varphi$, to $r+\rho$ for
$\theta=\varphi+\pi$, while $\theta\mapsto\theta'-\theta$ is odd under $\Gamma_\varphi$, increases
from 0 for $\theta=\varphi$ to a maximum value $(\theta'-\theta)_{\max}$, with
$\cos((\theta'-\theta)_{\max})= r(r^2+\rho^2)^{-1/2}$, for $\theta=\varphi+\pi/2$, and decreases
again to 0 for $\theta=\varphi+\pi$. Together with relation (b.10) this gives
$$|r'-r|/r\leq\rho/r,\quad |(\theta'-\theta)_{\max}|\approx\rho/r\leqno (b.12)$$
We compute (b.11) as follows. For $C>0$ large enough to be fixed later, we split the integral over
the sets $\{r\geq C \rho \}$, and $\{r\leq C \rho \}$. For the first part,
$(\theta'-\theta)_{\max}$ is a (non degenerate) critical point, and we observe that for $C>0$ large
enough~:
$$2\int_0^{2\pi}d\theta \bigl(1-\cos n(\theta-\theta')\bigr)=8\int_\varphi^{\varphi+\pi/2}d\theta \bigl(1-\cos n(\theta-\theta')\bigr)
=\pi n^2\bigl({\rho\over r}\bigr)^2+{\cal O}\bigl({\rho\over r}\bigr)^3\leqno(b.13)$$ where ${\cal
O}\bigl({\rho\over r}\bigr)^3$ is asymptotic, as $\rho/r \to 0$, to $c_n\bigl({\rho\over
r}\bigr)^3$, for some $c_n>0$, all $n$ ($c_n$ increases to $+\infty$ with $n$.~) This shows that
the corresponding integral contributes with a logarithmic singularity
$$\pi n^2\int_{\Omega_0}  d\rho \rho^3 \widehat J(\rho){dr\over r}, \quad
\Omega_0=\{r>r_0, r\geq C\rho\}\leqno(b.14)$$ which we substract from $2\int_\Omega d\theta
d\varphi rdr \rho d\rho \widehat J(\rho)\bigl(1-\cos n(\theta-\theta')\bigr)$. [Actually we replace
the integration over $r>r_0$ in (b.13) and (b.14) by $N>r>r_0$ and let then $N\to\infty$.] Since
the $\theta$-integral is clearly independent of $\varphi$, integrating over $\varphi$ gives an
additional factor of $2\pi$. The term ${\cal O}\bigl({\rho\over r}\bigr)^3$ in (b.13) then
contributes to a finite, and positive integral. Consider next the integral over $r\leq C\rho$.
Interchanging the $dr d\rho$ integrals (which is legitimate after the cutoff $r<N$,~) we are lead
to estimate
$$2\int_0^\infty rdr \int_{r/C}^\infty d\rho \rho \widehat J(\rho)\int_0^{2\pi}d\varphi\int_0^{2\pi}d\theta
\bigl(1-\cos n(\theta-\theta')\bigr)$$ There we use simply that the $d\varphi d\theta$ integral is
bounded and take advantage of the rapid decrease of $J$ to bound the $\rho$-integral by a negative
power of $r$ to make convergent the resulting $r$-integral. Thus we proved
\medskip
\noindent {\bf Lemma b.4}: With the notations above, if $m(x)=m_\beta e^{in\theta}$, let
$$\eqalign{
\ren &\bigl({1\over 4}\int dx\int dyJ(x-y)|m(x)-m(y)|^2\bigr)\cr &={1\over 4}\int dx\int
dyJ(x-y)|m(x)-m(y)|^2-2(\pi nm_\beta)^2\int_{\Omega_1}  d\rho \rho^3 \widehat J(\rho){dr\over r}\cr
}\leqno(b.15)$$ Then $0<c_n\leq\ren \bigl({1\over 4}\int dx\int
dyJ(x-y)|m(x)-m(y)|^2\bigr)<+\infty$.
\medskip

Now we turn to the more general case, which is sufficient for our purposes, where $m(x)$ belongs to
the $\langle e_n\rangle$-sector, and $|m(x)|\to m_\beta$ as $x\to\infty$. First we renormalize the
mean field free energy ${\cal F}^1(m)=\int dx f_\beta(m(x))$ for $|m|-m_\beta\in L^2({\bf R}^2)$~;
taking advantage of the fact that $f_\beta$ attains its minimum at $m_\beta$, we let simply
$$0\leq{\cal F}^1_{\ren }(m)=\int dx \bigl(f_\beta(m(x))-f_\beta(m_\beta)\bigr)<+\infty\leqno(b.16)$$
he first inequality because   . Next we pass to the interaction term ${\cal F}^0(m)$ as we did for
$m(x)=m_\beta e^{in\theta}$. So let $r_0>0$ and split the integral as in (b.9)~; as before we are
left with the integral over $\{|x|>r_0,|y|>r_0\}$. Assume also that $m$ is continuously
differentiable there. Use Taylor formula to rewrite $m(x)-m(y)=\int_0^1\nabla
m(ty+(1-t)x)dt\cdot(x-y)$, let $z=ty+(1-t)x=re^{i\theta},x-y=\rho e^{i\varphi}$. Expanding the
product, multiplying by $J(x-y)=\widehat J(\rho)$, computing the Jacobian, and integrating, we find
$$\eqalign{
{1\over 4}&\int_{|x|>r_0} dx\int_{L_x} dyJ(x-y)|m(x)-m(y)|^2=\int_0^1dt\int_0^1dt' \int_0^\infty
\widehat J(\rho)\rho^3d\rho\int_0^{2\pi} d\varphi\cr &\times\int_0^{2\pi} d\theta\int_{r_0}^\infty
rdr\bigl[{\partial m\over\partial r}(z) {\partial \overline m\over\partial
r'}(z')\cos(\varphi-\theta)\cos(\varphi-\theta')\cr &+{1\over r}{\partial m\over\partial
\theta}(z){\partial \overline m\over\partial r'}(z') \sin(\varphi-\theta)\cos(\varphi-\theta')+
{\partial  m\over\partial r}(z){1\over r'}{\partial \overline m\over\partial \theta'}(z')
\sin(\varphi-\theta')\cos(\varphi-\theta)\cr &+{1\over r}{\partial m\over\partial \theta}(z){1\over
r'}{\partial \overline m\over\partial \theta'}(z')
\sin(\varphi-\theta)\sin(\varphi-\theta')\bigr]\cr }\leqno(b.20)$$ where
$z'=t'y+(1-t')x=r'e^{i\theta'}$. Because $y\in L_x$, the segment $[x,y]$ lies outside $B_2(0,r_0)$
so $r,r'\geq r_0$ and all terms in the integral on the RHS of (b.20) are well defined. To start
with, we make as before a cut-off in the $r$-integral in the RHS, so that we can split it into
different terms. The domain of integration is denoted by $\widetilde\Omega=\{0\leq t,t'\leq 1,
r>r_0, \rho>0, \varphi,\theta\in[0,2\pi]\}$. The observations leading to (b.12) can be exactly
repeated, changing $x$ to $z$, $y$ to $z'$, and $\rho$ to $|t-t'|\rho$, now we get~:
$$|r'-r|/r\leq |t-t'|\rho/r,\ \cos((\theta'-\theta)_{\max})=
r(r^2+(t-t')^2\rho^2)^{-1/2}, \ |(\theta'-\theta)_{\max}|\approx|t-t'|\rho/r\leqno (b.21)$$ Making
use of the symmetry in $\theta,\theta'$, (b.20) is the sum of 3 integrals, whose corresponding
integrands write~:
$$\eqalign{
A(t,t',\rho,r,\varphi,&\theta)=u'(r)u'(r')\cos(\varphi-\theta)\cos(\varphi-\theta')\cos
n(\theta-\theta')\cr B(t,t',\rho,r,\varphi,&\theta)=2n[{u(r)\over
r}u'(r')\sin(\varphi-\theta)\cos(\varphi-\theta')\cr &+u'(r){u(r')\over
r'}\sin(\varphi-\theta')\cos(\varphi-\theta)]\sin n(\theta-\theta')\cr
D(t,t',\rho,r,\varphi,&\theta)=n^2{u(r)u(r')\over rr'}\sin(\varphi-\theta)\sin(\varphi-\theta')\cos
n(\theta-\theta')\cr }\leqno(b.23)$$ ($u'(r)$ denotes the $r$-derivative of $u$. ) We examine first
$D$, which we renormalize by setting $u(r)=m_\beta+v(r)$~; using (b.10) we find~:
$$D=n^2\bigl[\bigl({m_\beta\over r}\bigr)^2+{m_\beta\over r^2}(v(r)+v(r'))+r^{-2}v(r)v(r')\bigr]
\sin^2(\varphi-\theta)\cos n(\theta-\theta')\leqno(b.24)$$ Comparing the coefficients of
$m_\beta^2$ in (b.20) and (b.11), (b.13) shows that we must have, for $r\geq C\rho$~:
$$n^2\bigl({\rho\over r}\bigr)^2\int_0^1dt\int_0^1dt'\int_0^{2\pi}d\theta\sin^2(\varphi-\theta)\cos n(\theta-\theta')=
\pi n^2\bigl({\rho\over r}\bigr)^2+{\cal O}\bigl({\rho\over r}\bigr)^3\leqno(b.25)$$ Since the
$\theta$-integral is clearly independent of $\varphi$, integrating over $\varphi$ gives again an
additional factor of $2\pi$, so that $\int_{\widetilde\Omega}dtdt'd\varphi\widehat
J(\rho)\rho^3d\rho rdr n^2\bigl({m_\beta\over r}\bigr)^2 \sin^2(\varphi-\theta)\cos
n(\theta-\theta') $ contributes to (b.20) [after letting $N\to\infty$] with the logarithmic
singularity
$$2(\pi nm_\beta)^2\int_{\Omega_0}  d\rho \rho^3 \widehat J(\rho){dr\over r}, \quad
\Omega_0=\{r>r_0, r\geq C\rho\}$$  which we eventually substract from (b.20), and the term ${\cal
O}\bigl({\rho\over r}\bigr)^3$ in (b.25) then contributes with a finite, and positive integral.

Consider next the integral over $r_0<r\leq C\rho$. Interchanging the $dr d\rho$ integrals we are
lead to estimate
$$\int_{r_0}^\infty dr r\bigl({m_\beta\over r}\bigr)^2 \int_{r/C}^\infty d\rho \rho^3 \widehat J(\rho)
\int_0^{2\pi}d\varphi\int_0^{2\pi}d\theta\sin^2(\varphi-\theta)\cos n(\theta-\theta')$$ There
again, we take advantage of the rapid decrease of $\widehat J$ to bound the $\rho$-integral by a
negative power of $r$ to make convergent the resulting $r$-integral. Now we examine all other terms
contributing to (b.20), namely $A(t,t',\rho,r,\varphi,\theta)$, $B(t,t',\rho,r,\varphi,\theta)$,
and the part
$$D'(t,t',\rho,r,\varphi,\theta)=n^2\bigl[{m_\beta\over r^2}(v(r)+v(r'))+r^{-2}v(r)v(r')\bigr]
\sin^2(\varphi-\theta)\cos n(\theta-\theta')$$ which is left from $D$. Consider first $D'$, and
recall $v\in L^2({\bf R}^+;rdr)$. For the first term we use Cauchy-Schwarz inequality to get
$$|\int_{r_0}^\infty rdr {m_\beta\over r^2}v(r)\sin^2(\varphi-\theta)\cos n(\theta-\theta')|
\leq \bigl(\int_{r_0}^\infty rdr |v(r)|^2\bigr)^{1/2} \bigl(\int_{r_0}^\infty rdr
\bigl({m_\beta\over r^2}\bigr)^2\bigr)^{1/2}$$ and similarly for the second term, if we think of
the fact that $r$ and $r'$ play symmetric r\^oles. Integrating against $\widehat J(\rho)\rho^3$
with respect to $dtdt'd\varphi d\theta d\rho$ gives finite quantities. The last term in $D'$
contains the correlations $r^{-2}v(r)v(r')$. Again we split the integration according to $\{r\geq
C\rho\}$ and $\{r\leq C\rho\}$, and for the second part, use the rapid decrease of $\widehat J$.
For the first part, we use instead that the ``translations'' $(t,t',\varphi,\theta,\rho)\mapsto
\bigl( r\mapsto v(r')\bigr)$ are uniformly continuous in $L^2([C\rho,+\infty[;rdr)$ when $C\rho\leq
r_0$~; this gives
$$\eqalign{
|\int_{\widetilde \Omega, r>C\rho} &dtdt'd\varphi \widehat J(\rho)\rho^3d\rho rdr d\theta
r^{-2}v(r)v(r')\cr &\times\sin^2(\varphi-\theta)\cos n(\theta-\theta')|\leq \Const
\|v;L^2([r_0,+\infty[;rdr)\|^2\cr }$$ Consider then $A$ or $B$. These terms involve the derivatives
$r^{-1}v'(r')$, $r'^{-1}v'(r)$, and the correlations $v'(r)v'(r')$, $r^{-1}u(r)v'(r')$,
$r'^{-1}u(r')v'(r)$, so they can be treated as above, provided $r^\delta v'(r)$ is in $L^2(rdr)$
for some $\delta>0$. This holds true when $u(r)\sim m_\beta+{\cal O}(1/r)$, $r\to\infty$ and this
relation can be differentiated, e.g. if $m$ has the asymptotic properties of a symbol in $1/r$.
Summing up, we proved the~:
\medskip
\noindent {\bf Proposition b.5}: If $m(x)=u(r)e^{in\theta}$ is bounded, $u=m_\beta+v$, with
$$v\in{\cal W}=\{L^2([0,1];r^{-1}dr)\cap H^1({\bf R}^+;rdr), (\cdot)^{\delta} v'\in L^2([1,+\infty[;rdr)\}$$
for some $\delta>0$, then~:
$$\eqalign{
{\cal F}_{\ren ,r_0}(m)=&{1\over 4}\int dx\int dyJ(x-y)|m(x)-m(y)|^2-2(\pi
nm_\beta)^2\int_{\Omega_0} d\rho \rho^3 \widehat J(\rho){dr\over r}\cr &+\int dx
\bigl(f_\beta(m(x))-f_\beta(m_\beta)\bigr)<\infty\cr }\leqno(b.28)$$ Moreover, ${\cal F}_{\ren }$
is (strongly) continuous on ${\cal W}$ endowed with its natural Hilbert space structure.
\smallskip
\noindent {\it Remark b.6}: The renormalization can be easily extended to the case where $u$ has a
discontinuity on a sphere $r=\lambda$, i.e. when it has derivatives a.e. This is used when
considering the partial dynamics.
\smallskip
When $m$ is bounded, it is clear that ${\cal F}^0_{\ren ,r_0}(m)$ and ${\cal F}^0_{\ren ,r_1}(m)$
will just differ by a quantity of order $(r_0-r_1)^4$. Next we show that if $J\geq 0$, then ${\cal
F}_{\ren ,r_0}(m)$ is bounded from below in some region of the configuration space. First we
consider the interaction term ${\cal F}^0(m)={1\over 4}\int dx dyJ(x-y)|m(x)-m(y)|^2$, and compute
the formal Fourier transform of $h:{\bf R}^2\to{\bf C}$,
$h(x,y)=\sqrt{J(x-y)}\bigl(m(x)-m(y)\bigr)$. We have ${\cal F}h(\xi,\eta)=\bigl({\cal F}\sqrt
J(-\eta)-{\cal F}\sqrt J(\xi)\bigr){\cal F}m(\xi+\eta)$, so by Parseval identity (still in the
formal sense)
$$\int dxdyJ(x-y)|m(x)-m(y)|^2=(2\pi)^{-4}\int d\xi d\eta|{\cal F}m(\xi-\eta)|^2
\bigl({\cal F}\sqrt J(\eta)-{\cal F}\sqrt J(\xi)\bigr)^2$$ so we have exchanged the r\^oles of $m$
and $J$. Of course the integral is divergent because $m\notin L^2$, but the singularity is due to
$|{\cal F}m(\xi-\eta)|^2$, not to the boundary condition at infinity, since $\deg _\infty {\cal
F}\sqrt J=0$. The finite part (P.f.) of $|{\cal F}m(\xi-\eta)|^2$ is $|{\cal
F}m(\xi-\eta)|^2|_{\xi=\eta}=\bigl(\int mdx)^2$ and is finite when $m(x)=e^{in\theta}u(r)$,
$u(r)=m_\beta+v(r)$, $v\in L^2$, because of the periodicity in $\theta$. So another renormalization
of ${\cal F}^0(m)$ is given by
$${\cal F}^0_{\ren ,{\cal F}}(m)={1\over 4}(2\pi)^{-4} \ \hbox{P.f.} \ \int d\xi d\eta|{\cal F}m(\xi-\eta)|^2
\bigl({\cal F}\sqrt J(\eta)-{\cal F}\sqrt J(\xi)\bigr)^2$$ which is obviously positive. Moreover,
Lemma b.4 and its proof show that both renormalizations should agree up to a finite term, depending
only on $r_0$, when $m(x)=m_\beta e^{in\theta}$. Thus, we should have ${\cal F}^0_{\ren
,r_0}(m)\equiv {\cal F}^0_{\ren ,{\cal F}}(m)$ modulo a constant term, depending on $r_0$, but not
on $m$, when $m$ satisfies all hypotheses of Proposition b.5. In particular, ${\cal F}^0_{\ren
,r_0}(m)$ is bounded from below.

\bigskip
\noindent{\bf C). The free energy in Kac's model with continuous symmetry}.
\smallskip
In this  Section,  we recall and make more precise the procedure of renormalization in the
continuum, carried out in [AlBeCaPr], [Pr] for spins valued in $\{-1,+1\}$, and extended to the XY
model in [BuPi],  when a vorticity condition holds at infinity and the interaction is  not
necessarily compactly supported. Actually, the  final form for the continuous renormalized free
energy functional should be regarded as a postulate.

Note that passing from the lattice to the continuum amounts to consider convergence of Riemann sums
as the mesh of the (scaled) lattice goes to 0, as an homogenization process, i.e. the convergence
of discrete measures.
\medskip
\noindent {\bf a) Some definitions}.
\smallskip
Consider the lattice ${\bf Z}^d$, consisting in a bounded, connected domain $\widetilde\Lambda$
(the interior region), and its complement (the exterior region) $\widetilde\Lambda^c$ (twiddled
letters will always denote discrete objects on ${\bf Z}^d$.~) Physical systems make sense in the
thermodynamical limit $\widetilde\Lambda\to{\bf Z}^d$, in the sense of Fisher. The simplest way of
taking this limit is to double the side of the unit hypercube repeatedly, so the side of
$\widetilde\Lambda=\widetilde\Lambda_{\widetilde\ell}$ is of the form $2^{\widetilde\ell},
\widetilde\ell\in{\bf N}$.

To each site $i\in{\bf Z}^d$ is attached a classical spin variable $\sigma(i)\in{\bf S}^q$. The
configuration space ${\cal X}({\bf Z}^d)=({\bf S}^{q-1})^{{\bf Z}^d}$ is the set of all such
classical states of spin~; it has the natural internal symmetry group $O^+(q)$ ($q=2$ for the
planar rotator.) Given the partition ${\bf Z}^d=\widetilde\Lambda\cup\widetilde\Lambda^c$, we
define by restriction the interior and exterior configuration spaces ${\cal X}(\widetilde\Lambda)$
and ${\cal X}(\widetilde\Lambda^c)$, and the restricted  configurations by
$\sigma_{\widetilde\Lambda}$ and $\sigma_{\widetilde\Lambda^c}$.

It is convenient to rescale $\widetilde\Lambda_\ell$ to a domain of fixed size $L=2^{\ell}$,
$\ell\in{\bf N}$, $\Lambda\subset{\bf R}^d$, with scaling factor is $2^{-\widetilde\ell}$~; for the
moment we could think of $\Lambda$ as the unit square ($\ell=0$) but we shall eventually let also
$\Lambda\to\infty$.

Following [Pr], for $k\in{\bf N}$, we denote by ${\cal Q}^{(k)}$ the partition  of ${\bf R}^d$ into
small cubes $C^{(k)}=\{r=(r_1,\cdots,r_d)\in{\bf R}^d, 2^{-k}x_i\leq r_i<2^{-k}(x_i+1)\}$, of side
$2^{-k}$, and indexed by $x=(x_1,\cdots,x_d)\in{\bf Z}^d$, called (${\cal Q}^{(k)}$-)atoms. The
atom $C^{(k)}(r)$ is the unique atom of ${\cal Q}^{(k)}$ that contains $r$. We say also that a
function on ${\bf R}^d$ is ${\cal Q}^{(k)}$-measurable if it is constant on each atom of  ${\cal
Q}^{(k)}$, and a set $A\subset{\bf R}^d$ is ${\cal Q}^{(k)}$-measurable if its indicator function
is ${\cal Q}^{(k)}$-measurable. This allows in a natural way to identify a function $\sigma$ on the
lattice  with a ${\cal Q}^{(k)}$-measurable function $\sigma^{(k)}$ on ${\bf R}^d$, assuming
$\sigma^{(k)}(r)=\sigma(x)$ with $x=(x_1,\cdots,x_d)$ and $r=(r_1,\cdots,r_d)$ as above.

Let now $\gamma$ of the form  $\gamma=2^{-k_\gamma}$, $k_\gamma\in{\bf N}$, that will be the
inverse of the interaction length in Kac's potential. Given  a state  $\sigma\in {\cal X}({\bf
Z}^d)$, we define $\sigma_\gamma$ as the ${\cal Q}^{(k_\gamma)}$-measurable function
$\sigma^{(k_\gamma)}$. Since we take a simultaneous limit $\widetilde\Lambda\to\infty$, $\gamma\to
0$, rather than Lebowitz-Penrose limit $\widetilde\Lambda\to\infty$  followed  by $\gamma\to 0$, it
may be convenient to label the configurations by $\gamma$, instead of $\Lambda$. We call also
$\sigma_\gamma$ a {\it smooth-grained} configuration.

Because of the internal continuous symmetry of ${\cal X}(\Lambda)$, the probability distribution
$\nu$ for the states of spin is defined as the normalized surface measure on ${\bf S}^{q-1}$, i.e.
$\nu(d\sigma_i)=\omega_q^{-1}\delta(|\sigma_i|-1)d\sigma_i$, where $\omega_q$ is the volume of
${\bf S}^{q-1}$.

Given $\sigma_\gamma$ as above, and an integer $n_\gamma\leq k_\gamma$, we associate the ${\cal
Q}^{(n_\gamma)}$-measurable function (magnetization)
$$m_\gamma(r)=\pi^{(n_\gamma)}\sigma_\gamma(r)={1\over|C^{(n_\gamma)}|}\int_{C^{(n_\gamma)}(r)}dr'\sigma_\gamma(r')
\leqno(c.1)$$ These averages of $\sigma_\gamma$ over the ``intermediate'' boxes (or block spins)
$C^{(n_\gamma)}(r)$ of volume $\gamma^{\delta d}$, define the {\it coarse-grained} configurations,
and the map $\pi^{(n_\gamma)}:\Omega\to B'_q(0,1)$, $\Omega=\bigl({\bf S}^{q-1}\bigr)^{\widetilde
C^{(n_\gamma)}},B'_q(0,1)$ the closed unit ball of ${\bf R}^q$, is called the {\it block-spin
transformation}. We may think of it as a random variable.

More generally, let $\pi_N:\bigl({\bf S}^{q-1}\bigr)^N\to
B'_q(0,1),\sigma=(\sigma_1,\cdots,\sigma_N)\mapsto\pi_N(\sigma)= {1\over N}\Sum_{i=1}^N\sigma_i$.
It is easy to see that $\pi_N$ is a smooth, surjective map, and its restriction
$\widetilde\pi_N:\bigl({\bf S}^{q-1}\bigr)^N\setminus\Delta\to B_q(0,1)$ a submersion, where
$\Delta$ denotes the diagonal $\sigma_1=\cdots=\sigma_N$ of $\bigl({\bf S}^{q-1}\bigr)^N$, and
$B_q(0,1)$ the open unit ball of ${\bf R}^q$. Hence the probability measure
$\nu_N=(\pi_N)_*(\nu\otimes\cdots\otimes\nu)$ has a smooth density with respect to the (normalized)
Lebesgue measure on $B_q(0,1)$, namely ${d\nu_N\over dm}(m)=\int_{\bigl({\bf S}^{q-1}\bigr)^N}
\prod_{i=1}^N\nu(d\sigma(i))\delta\bigl(\pi_N(\sigma)-m\bigr)$, and we can check as in
[BuPi,formula (5.1)] that
$${d\nu_N\over dm}(m)=\bigl({N\over 2\pi}\bigr)^q
\int_{{\bf R}^q}dve^{-iN\langle v,m\rangle}\bigl(\int_{{\bf S}^{q-1}}\nu(dx)e^{i\langle
v,x\rangle}\bigr)^N$$ (here $\langle\cdot,\cdot\rangle$ is the standard scalar product in ${\bf
R}^q$.~)

Of course, Kolmogorov realization theorem would allow to define this way a probability measure on
$\bigl(B_q(0,1)\bigr)^\Lambda$, but this will not be used at the present level.

Now, following [Pr] we choose  $n_\gamma$ so that $\gamma^\delta=2^{-n_\gamma}$, for some
$0<\delta<1$, a good choice is $\delta=1/2$. From the point of view of  the  discrete scheme, a
coarse configuration is defined on a ``coarse lattice'' $\widetilde\Lambda^*$ which we magnify by
the factor $2^{k_\gamma-n_\gamma}$ to the ``smooth lattice'' $\widetilde\Lambda$ (see e.g.
[El-BoRo].~) Thermodynamical properties of the system are most significant on the coarse lattice.

From the discussion above, for $k\in{\bf N}$, let $\widetilde C^{(k)}=2^kC^{(k)}\cap{\bf Z}^d$
denote the atom $C^{(k)}$ rescaled to the lattice units~; thus, $\widetilde C^{(k)}$ is the
rescaled block-spin. So the probability distribution $\nu^{(n_\gamma)}$ of the empirical average
$\pi^{(n_\gamma)}$ has Radon-Nikodym density
$${d\nu^{(n_\gamma)}\over dm}(m)=
\int_\Omega \prod_{i\in\widetilde
C^{(n_\gamma)}}\nu\bigl(d\sigma_{\widetilde\Lambda}(i)\bigr)\delta\bigl(\pi^{(n_\gamma)}
\sigma_{\widetilde\Lambda}(i)-m\bigr),\quad\Omega=\bigl({\bf S}^{q-1}\bigr)^{\widetilde
C^{(n_\gamma)}} \leqno(c.2)$$ We shall sometimes write $\sigma_\gamma(i)$ instead of
$\sigma_{\widetilde\Lambda}(i)$. Let also $N=|\widetilde C^{(n_\gamma)}|=\gamma^{(\delta-1) d}$,
$\nu^{(n_\gamma)}=\nu_N$.

The continuous Kac Hamiltonian is defined as follows.  Let $J\geq 0$ be the interaction potential,
for a given  $\sigma^c\in{\cal X}(\widetilde\Lambda^c)$, and for  $\Lambda=\bigcup
C^{(k_\gamma)}=\bigcup C^{(n_\gamma)}$ as above, we define the energy of a configuration
$\sigma\in{\cal X}(\widetilde\Lambda)$ by
$$H(\sigma_\gamma|\sigma^c_\gamma)=-{1\over 2}\int_\Lambda dr\int_\Lambda dr'
J(r-r')\langle\sigma_\gamma(r),\sigma_\gamma(r')\rangle -\int_\Lambda dr\int_{\Lambda^c} dr'
J(r-r')\langle\sigma_\gamma(r),\sigma^c_\gamma(r')\rangle \leqno(c.3)$$ Again from the point of
view of  the  discrete scheme, since $\sigma_\gamma(\gamma i)=\sigma(i)$, for $i\in{\bf Z}^d$,
there is a corresponding Hamiltonian on ${\bf Z}^d$ defined by
$$\widetilde H_\gamma(\sigma_{\widetilde\Lambda}|\sigma_{\widetilde\Lambda^c})
=-{1\over 2}\Sum_{i,j\in\widetilde\Lambda}J_\gamma(i,j)\langle\sigma(i), \sigma(j)\rangle-
\Sum_{(i,j)\in\widetilde\Lambda\times\widetilde\Lambda^c} J_\gamma(i,j)\langle\sigma(i),
\sigma(j)\rangle \leqno(c.4)$$ where
$$J_\gamma(i,j)=\gamma^{-d}\int_{C^{(n_\gamma)}(\gamma i)} dr\int_{C^{(n_\gamma)}(\gamma j)} dr' J(r-r')\leqno(c.5)$$
The two hamiltonians are simply related by~:
$$\widetilde H_\gamma(\sigma_{\widetilde\Lambda}|\sigma_{\widetilde\Lambda^c})=\gamma^{-d}
H(\sigma_\gamma|\sigma^c_\gamma)\leqno(c.6)$$ so that $H$ is an intensive hamiltonian. As observed
in [AlBeCaPr], neglecting the variations of $J$ in the integral, we get
$$J_\gamma(i,j)\approx\gamma^{d}J(\gamma|i-j|)\leqno(c.7)$$
which has the typical scaling properties  of the original Kac potential, and the results of this
Section remain valid when the energy is given by (c.4) with (c.7) holding as  an equality. This
observation allows to define Gibbs measure conditioned by $\sigma_{\widetilde\Lambda^c}$, on the
space of spin configurations ${\cal X}(\widetilde\Lambda)$ with mesh $\gamma$, at inverse
temperature  $\beta$, as
$$\mu_{\beta,\gamma,\widetilde\Lambda}(d\sigma_{\widetilde\Lambda}|\sigma_{\widetilde\Lambda^c})=
{1\over Z_{\beta,\gamma,\widetilde\Lambda}(\sigma_{\widetilde\Lambda^c})} \exp
\bigl[-\beta\widetilde H_\gamma(\sigma_{\widetilde\Lambda}|\sigma_{\widetilde\Lambda^c})\bigr]
\prod_{i\in\widetilde\Lambda}\nu\bigl(d\sigma_{\widetilde\Lambda}(i)\bigr)\leqno(c.8)$$ where
$$Z_{\beta,\gamma,\widetilde\Lambda}(\sigma_{\widetilde\Lambda^c})=
\int_{\Omega_0} \prod_{i\in\widetilde\Lambda}\nu\bigl(d\sigma_{\widetilde\Lambda}(i)\bigr) \exp
\bigl[-\beta\widetilde H_\gamma(\sigma_{\widetilde\Lambda}|\sigma_{\widetilde\Lambda^c})\bigr],
\quad \Omega_0=\bigl({\bf S}^{q-1}\bigr)^{\widetilde\Lambda} \leqno(c.9)$$ is the partition
function, making of
$\mu_{\beta,\gamma,\widetilde\Lambda}(\sigma_{\widetilde\Lambda}|\sigma_{\widetilde\Lambda^c})$ a
probability measure on the product space. As before, we can take the direct image
$\mu_{\beta,\gamma,\widetilde\Lambda}(d\sigma_{\widetilde\Lambda}|\sigma_{\widetilde\Lambda^c})$
through the block-spin transformation. We shall discuss this in the following subsections.

\medskip
\noindent {\bf b) Entropy estimates}.
\smallskip
We want to relate ${1\over N}\log {d\nu_N\over dm}(m)$ with the entropy functional $I(m)$ of the
mean field approximation. Recall $I(m)=\sup _{k\in{\bf R}^q}\bigl(\langle k,m\rangle-\log
\phi(k)\bigr)$, $\phi(k)=\int_{{\bf S}^{q-1}}e^{\langle k,v\rangle} \nu(dv)$, and $I(m)=\langle
k^*,m\rangle-\log \phi(k^*)$ where $k^*=k^*(m)$ is the unique point in ${\bf R}^q$ that achieves
the maximum. Clearly also, by spherical symmetry, $I(m)=\widehat  I(|m|)$, and $\widehat
I(|m|)=\sup _{t\geq 0}\bigl(t|m|-\log \widehat\phi(t)\bigr)=t^*|m|-\log \widehat\phi(t^*)$, where
$t^*=t^*(|m|)$, $t^*(0)=0, t^*(\rho)\sim(2-2\rho)^{-1}$ as $\rho\to 1$. Furthermore the supremum is
achieved when $k^*$ and  $m$ are colinear. For $|m|<1$ we introduce the probability measure
$\mu(dx;m)$ on ${\bf S}^{q-1}$ defined by
$$\mu(dx;m)=\exp(\langle k^*,x\rangle-\log \phi(k^*)) \nu(dx)\leqno(c.11)$$
As in (c.2) we define
$${d\mu^{(n_\gamma)}\over dm}(m)=
\int_\Omega \prod_{i\in\widetilde
C^{(n_\gamma)}}\mu\bigl(d\sigma_{\widetilde\Lambda}(i);m\bigr)\delta\bigl(\pi^{(n_\gamma)}
\sigma_{\widetilde\Lambda}(i)-m\bigr)$$ and denote $\mu_N(dx;m)=\mu^{(n_\gamma)}(dx;m)$. It is easy
to see that ${d\nu_N\over dm}(m)=e^{-NI(m)}{d\mu_N\over dm}(m)$. Let also $\varphi_m$ be the
complex function  defined  on ${\bf R}^q$
$$\varphi_m(v)=e^{i\langle v,m\rangle} \int_{{\bf S}^{q-1}}\mu(dx;m)e^{-i\langle v,x\rangle}$$
Recall from [BuPi,formula(5.7)] the identity
$${1\over N}\log {d\nu_N\over dm}(m)+I(m)={1\over N}\log\bigl[\bigl({N\over 2\pi}\bigr)^q
\int_{{\bf R}^q}dv\varphi_m(v)^N\bigr]\leqno(c.12)$$ The observation is that ${1\over N}\log
{d\nu_N\over dm}(m)$ is a small correction to $-I(m)$ (the entropy for the mean field) as $N$
becomes large. Indeed we have~:
\medskip
\noindent{\bf Lemma c.1}: Let $q=2$ for simplicity. With the notations above, we have for all
$N\geq 1$,
$$|{1\over N}\log {d\nu_N\over dm}(m)+I(m)|\leq{1\over N}\log [C_0\bigl({N\over 2\pi}\bigr)^q(N^q+N^{q'}(1-|m|)^{-1/2})]
\leqno(c.13)$$ for some $C_0>0,q'\geq 0$.
\smallskip
\noindent {\it Proof}: We need to show that the integral on the RHS of (c.12) grows at most
linearly in $(1-|m|)^{-1/2}$, with coefficients polynomial in $N$. Notice first that
$|\varphi_m(v)|\leq\varphi_m(0)=1$ for all $v\in{\bf R}^q$, so integrating $\varphi_m(v)^N$ over
the ball in ${\bf R}^q$ of center 0 and radius $N$ we get
$$|\int_{|v|\leq N}dv \varphi_m(v)^N|\leq \Const N^q\leqno(c.15)$$
Now, we estimate the integral near $\infty$, using complex stationary phase. We will be a  little
sketchy, but it is easy to see that our leading terms give the correct behavior with the required
uniformities (see e.g. [Sj] for more details.~) Let $v=r(\cos\varphi,\sin\varphi)$,
$\varphi\in[-\pi,\pi]$, we rewrite $\varphi_m(v)$ as
$$\varphi_m(v)=\bigl(2\pi\widehat\phi(t^*)\bigr)^{-1}e^{ir|m|\sin\varphi}\int_{-\pi}^\pi d\theta e^{-ir\Phi(\theta,\varphi)},
\quad  \widehat\phi(t^*)=I_0(t^*) \leqno(c.16)$$ with
$\Phi(\theta,\varphi)=\cos(\theta-\varphi)+i\lambda\sin\theta$, and $\lambda=t^*/r$. Here we
consider $r\geq N$ as the large parameter. The critical points in $\theta$ are given by the
equation $\sin(\theta-\varphi)-i\lambda\cos\theta=0$, so $\theta\mapsto\Phi(\theta,\varphi)$  has
no real critical point if $\varphi\neq\pm\pi/2$, and 2 real critical points $\theta=\pm\varphi$
otherwise. Actually, $\im \Phi(\pm\pi/2,\mp\pi/2)<0$ so the contribution of the critical point with
sign opposite to this of $\varphi$ will be exponentially small, and by symmetry it suffices to
consider $(\theta,\varphi)=(\pi/2,\pi/2)$. This is a non degenerate critical point, since
${\partial^2\Phi\over\partial\theta^2}(\pi/2,\pi/2)=-1-i\lambda$. Because of analyticity, there is
a complex critical point $\theta_c=\theta_c(\varphi)$ for nearby values of $\varphi$, and a simple
calculation yields
$$\Phi(\theta_c,\varphi)=1+i\lambda-{i\lambda\over 2(1+i\lambda)}(\varphi-\pi/2)^2+{\cal O}(\varphi-\pi/2)^3\leqno(c.17)$$
where ${\cal O}(\varphi-\pi/2)^3$ is uniform in $\lambda$. The complex Morse lemma then shows that
the local  analytic diffeomorphism $\theta\mapsto\widetilde\theta$ given by
$\widetilde\theta=\sqrt{f_1(\theta-\theta_c;\varphi)}e^{-i\pi/4}(\theta-\theta_c)$,
$f_1(0;\pi/2)=1+i\lambda$, is such that
$\Phi(\theta,\varphi)=\Phi(\theta_c,\varphi)-i\widetilde\theta^2/2$. Then complex stationary phase
shows that the contribution of a (fixed) neighborhood of $\theta_c$ to the integral in (c.16) is
given at leading order, by $\bigl({2\pi\over
rf_1(0;\varphi)}\bigr)^{1/2}e^{i\pi/4}e^{-ir\Phi(\theta_c,\varphi)}$. Outside this neighborhood,
non stationary phase arguments show that the integral is exponentially smaller than
$e^{-ir\Phi(\theta_c,\varphi)}$, as a function of $r$, and we eventually get, for $\varphi$ in a
(real) neighborhood $V_\pm$ of $\pm\pi/2$~:
$$\int_{-\pi}^\pi d\theta e^{-ir\Phi(\theta,\varphi)}=
\bigl({2\pi\over
rf_1(0;\varphi)}\bigr)^{1/2}e^{i\pi/4}e^{-ir\Phi(\theta_c,\varphi)}(1+R_1(\varphi,\lambda,r)/r)
\leqno(c.18)$$ with $|R_1(\varphi,\lambda,r)|\leq \Const $. By analyticity, we can keep track of
the critical point $\theta_c$ for all $\varphi\in [-\pi,\pi]$, so formula (c.18) still makes sense
when $\varphi\notin V_\pm$, but then $\im \Phi(\theta_c,\varphi)<t^*$, and $\varphi\notin V_\pm$
doesn't contribute to the final result. Now we raise (c.16) to the power $N$ and integrate  over
$r\geq N$, using (c.18) this yields (with a factor 2, accounting for the contribution $\varphi\in
V_-$~ )
$$\eqalign{
\int_{|v|\geq r_1}dv &\varphi_m(v)^N\sim 2(2\pi\widehat\phi(t^*))^{-N}e^{iN\pi/4}(2\pi)^{N/2}
\int_{r\geq N}rdr(rf_1(0;\varphi))^{-N/2}\cr &\times\int_{V_+} d\varphi
e^{-iNr\Phi_1(\varphi)}(1+R_1(\varphi,\lambda,r)/r)^N\cr }\leqno(c.19)$$ where
$\Phi_1(\varphi)=\Phi(\theta_c,\varphi)-|m|\sin\varphi$.  Then (c.17) shows that $\Phi_1$ has a non
degenerate point at $\varphi=\pi/2$, and as before, the complex Morse lemma gives a local analytic
diffeomorphism $\varphi\mapsto\widetilde\varphi=f_2(\varphi-\pi/2)(\varphi-\pi/2)$ with
${d\widetilde\varphi\over d\varphi}(\pi/2)=f_2(0)=\bigl({\lambda\over
1+i\lambda}+i|m|\bigr)^{1/2}$, such that $\Phi_1(\varphi)-\Phi_1(\pi/2)={1\over
2}\widetilde\varphi^2$, $\Phi_1(\pi/2)=1-|m|+i\lambda$. So by complex stationary phase, with $Nr$
as a large  parameter, we can evaluate the inner integral in (c.19)~:
$$\eqalign{
\int_{V_+} d\varphi &e^{-iNr\Phi_1(\varphi)}(1+R_1(\varphi,\lambda,r)/r)^N=\cr
&e^{-iNr(1-|m|)}e^{Nt^*}\sqrt{2\pi/Nr}f_2(0)^{-1}(1+R_1(\pi/2,\lambda,r)/r)^N(1+R_2(\lambda,|m|,r)/Nr)\cr
}\leqno(c.20)$$ with $R_2(\lambda,|m|,r)\leq \Const $. At last, we estimate the resulting
$r$-integral in (c.19). This time we compute simply an upper bound for the integrand. We have
$(1+R_1(\pi/2,\lambda,r)/r)^N\leq\Const $ uniformly in $N$ as $r\geq N$, and the same holds for
$(1+R_2(\lambda,|m|,r)/Nr)$, so inserting (c.20) into (1.19), we get
$$\int_{|v|\geq r_1}dv \varphi_m(v)^N\leq \Const \int_{r\geq N}rdr
\bigl({e^{t^*}\over  \sqrt{2\pi}I_0(t^*)}|r+it^*|^{-1/2}\bigr)^N \sqrt{2\pi/Nr}|f_2(0)|^{-1}
\leqno(c.21)$$ We have
$$\bigl({e^{t^*}\over\sqrt{2\pi}I_0(t^*)}|r+it^*|^{-1/2}\bigr)^N=
\bigl({e^{t^*}\over\sqrt{2\pi}I_0(t^*)}(N^2+t^{*2})^{-1/4}\bigr)^N\bigl({N^2+t^{*2}\over
r^2+t^{*2}}\bigr)^{N/4}$$ and since ${e^{t^*}\over\sqrt{2\pi t^*}I_0(t^*)}\to 1^-$ as
$t^*\to+\infty$, it is easy to see that
$$\limsup _{N\to\infty}\bigl({e^{t^*}\over \sqrt{2\pi}I_0(t^*)}(N^2+t^{*2})^{-1/4}\bigr)^N\leq\Const $$
uniformly in $t^*>0$. Furthermore $|f_2(0)|^{-1}=({t^{*2}+r^2)^{1/4}
(t^{*2}(1-|m|)^2+r^2|m|^2}\bigr)^{-1/4}$, and since $t^*\sim 2^{-1}(1-|m|)^{-1}$ as $|m|\to 1$, we
have $|f_2(0)|^{-1}\leq\Const (1-|m|)^{-1/2}$ uniformly as $r\geq N$. So the integral on the RHS of
(c.21) is bounded by a constant times
$$(1-|m|)^{-1/2}\int_{r\geq N}rdr\sqrt{2\pi/Nr}\bigl({N^2+t^{*2}\over r^2+t^{*2}}\bigr)^{N/4}$$
and it is easy to show that there is $q'\geq0$ such that $\int_{r\geq
N}rdr\sqrt{2\pi/Nr}\bigl({N^2+t^{*2}\over r^2+t^{*2}}\bigr)^{N/4}\leq\Const N^{q'}$ uniformly in
$t^*>0$. This completes the proof of the Lemmc. $\clubsuit$
\smallskip
Next we have the following~:
\medskip
\noindent{\bf Proposition c.2}: With the notations above, there is $C_1>0$ such that
$$|\log {d\nu^{(\Lambda)}\over dm}+\gamma^{-d}I(m,\Lambda)|\leq
C_1(L\gamma^{-\delta})^d\log\gamma^{-1}+\log\prod_{x\in\widetilde \Lambda/\widetilde
C^{(n_\gamma)}}\bigl(1-|m(x)|\bigr)^{-1/2} \leqno(c.25)$$ \noindent{\it Proof}: Sum (c.13) over all
the cubes $\widetilde C^{(n_\gamma)}$ contained in $\widetilde\Lambda$, which have same cardinal
$N=\gamma^{(\delta-1)d}$, and multiply by $N$ the resulting equality. The first term on the LHS
$$\log\prod_{\widetilde C^{(n_\gamma)}} {d\nu^{(n_\gamma)}\over
dm}(m)=\int_{\Omega_0} \prod_{i\in\widetilde
\Lambda}\nu\bigl(d\sigma_{\widetilde\Lambda}(i)\bigr)\delta\bigl(\pi^{(n_\gamma)}
\sigma_{\widetilde\Lambda}(i)-m\bigr)\leqno(c.26)
$$ is the Radon-Nikodym density ${d\nu^{(\Lambda)}\over dm}(m)$
of a probability distribution, namely the family of empirical averages $\pi^{(n_\gamma)}$ in
$\widetilde C^{(n_\gamma)}$ considered as i.i.d. random variables. Here we need interpret $m$ as a
${\cal Q}^{(n_\gamma)}$-measurable function on ${\bf R}^d$, and the RHS of (c.24) should actually
read
$$\int_{\Omega_0}\prod_{x\in\widetilde \Lambda/\widetilde C^{(n_\gamma)}}\prod_{i\in
\widetilde
C^{(n_\gamma)}_x}\nu\bigl(d\sigma_{\widetilde\Lambda}(i)\bigr)\delta\bigl(\pi^{(n_\gamma)}
\sigma_{\widetilde\Lambda}(i)-m(x)\bigr)$$ where somewhat incorrectly, the notation $\widetilde
\Lambda/\widetilde C^{(n_\gamma)}$ reminds us we have tiled $\widetilde\Lambda$ by translates of
$\widetilde C^{(n_\gamma)}$, and $\widetilde C^{(n_\gamma)}_x$ is the atom of the partition of
$\widetilde\Lambda$ labelled by $x$. We have $|\widetilde \Lambda/\widetilde
C^{(n_\gamma)}|=|\Lambda/ C^{(n_\gamma)}|=(L\gamma^{-\delta})^d$. This identification will be made
freely in the sequel. Summing up the second terms on the LHS of (c.13) over $x$ will produce
$\gamma^{-d}I(m,\Lambda)$, where $I(m,\Lambda)=\int_\Lambda drI(m(r))$ defines a functional on the
space of ${\cal Q}^{(n_\gamma)}$-measurable functions. While summing up the RHS of (c.13) over $x$,
(c.25) follows easily if we make use of the inequality $\log(a+b)\leq\log(1+a)+\log b$ valid for
any $a>0, b\geq 1$. $\clubsuit$
\medskip
\noindent {\bf c) Free energy estimates}.
\smallskip
Following [Pr], we replace now in the hamiltonian the spins $\sigma_\gamma(r)$ by the magnetization
$m_\gamma(r)$ as in (c.1). We have~:
\medskip
\noindent {\bf Lemma c.3}: With the notations above, for some $C_2>0$ we have :
$$|H(m_\gamma|m^c_\gamma)-H(\sigma_\gamma|\sigma^c_\gamma)|\leq C_2 L^d \gamma^\delta\|\nabla J\|_1
\leqno(c.30))$$ (here $\|\cdot\|_1$ denotes the $L^1$-norm on ${\bf R}^d$)
\smallskip
\noindent {\it Proof}: By the definition of $H(\sigma_\gamma|\sigma^c_\gamma)$ and (c.1) we have,
denoting $C_\gamma(r)$ for $\widetilde C^{(n_\gamma)}(r)$
$$\eqalign{
&H(m_\gamma|m^c_\gamma)-H(\sigma_\gamma|\sigma^c_\gamma)=\bigl({1\over 2}\int_\Lambda
dr_1\int_\Lambda dr_2+ \int_\Lambda dr_1\int_{\Lambda^c} dr_2\bigr)
\langle\sigma_\gamma(r_1),\sigma_\gamma(r_2)\rangle\cr &\times\bigl[\gamma^{-2d\delta}
\int_{C_\gamma(r_1)}dr\int_{C_\gamma(r_2)}dr'\bigl(J(r-r')-J(r_1-r_2)\bigr)\bigr]\cr }$$ where we
have used $r_1\in C_\gamma(r)$ iff $r\in C_\gamma(r_1)$. We estimate $J(r-r')-J(r_1-r_2)$ by Taylor
formula, noticing  that $|(r-r')-(r_1-r_2)|\leq \Const \gamma^\delta$, so the $dr_2$ integrals over
$\Lambda$ or $\Lambda^c$ are bounded by a constant times $\gamma^\delta\|\nabla J\|_1$, then the
resulting $dr_1$ integral over $\Lambda$ gives an additional $L^d$ factor, which proves the Lemma.
$\clubsuit$.
\smallskip
Introduce the free energy in $\Lambda$ at inverse temperature $\beta$, inclusive of the interaction
on $\Lambda^c$, as the functional on the space of ${\cal Q}^{(n_\gamma)}$-measurable functions
$m(r)$
$$\eqalign{
{\cal F}(m|m^c)&={1\over 4}\int_\Lambda dr\int_\Lambda dr'J(r-r')|m(r)-m(r')|^2\cr &+{1\over
2}\int_\Lambda dr\int_{\Lambda^c} dr'J(r-r')|m(r)-m(r')|^2+ \int_\Lambda dr
\bigl(f_\beta(m(r))-f_\beta(m_\beta)\bigr)\cr }\leqno(c.31)$$ where we recall $f_\beta(m)=-{1\over
2}|m|^2+{1\over\beta}I(m)$. Let $\widehat e$ be any (fixed) unit vector in ${\bf R}^q$, and
$\widehat m_\beta$ the constant function on $\Lambda$ equal to $m_\beta\widehat e$, which we extend
to be equal to $m^c$ on $\Lambda^c$. The functionals $I(\cdot,\Lambda)$ and $H(\cdot|m^c)$ are
related to  ${\cal F}(\cdot|m^c)$ in a simple way~:
$${\cal F}(m|m^c)-{\cal F}(\widehat m_\beta|m^c)
=(H(m|m^c)+{1\over\beta} I(m,\Lambda))-(H(\widehat m_\beta|m^c)+{1\over\beta} I(\widehat
m_\beta,\Lambda))\leqno(c.32)$$ Analogously to (c.8), (c.2) we introduce the canonical Gibbs
measure conditioned by the external configuration $\sigma_{\widetilde\Lambda^c}=\sigma_\gamma^c$~:
$$\mu_{\beta,\gamma,\Lambda}(d\sigma_\gamma;m|\sigma_\gamma^c)=
{1\over  Z_{\beta,\gamma,\Lambda}(\sigma_\gamma^c)}\int_{\Omega_0} \exp
\bigl[-\beta\gamma^{-d}H_\gamma(\sigma_\gamma|\sigma_\gamma^c)\bigr]
\prod_{i\in\widetilde\Lambda}\nu\bigl(d\sigma_{\widetilde\Lambda}(i)\bigr)\delta(\pi_\gamma\sigma_\gamma(i)-m)\leqno(c.33)$$
where the  partition function
$Z_{\beta,\gamma,\Lambda}(\sigma_\gamma^c)=Z_{\beta,\gamma,\widetilde\Lambda}(\sigma_{\widetilde\Lambda^c})$
was defined  in (c.9). We have made use of (c.5) to work on  the rescaled lattice  $\Lambda$,  and
set
$\delta(\pi_\gamma\sigma_\gamma(i)-m)=\delta\bigl(\pi^{(n_\gamma)}\sigma_{\widetilde\Lambda}(i)-m\bigr)$.
By definition of the image of Gibbs measure through the block-spin transformation, we have
$$\int_{|m|<1}dm\mu_{\beta,\gamma,\Lambda}(d\sigma_\gamma;m|\sigma_\gamma^c)=1\leqno(c.34)$$
where $dm$ is the normalized Lebesgue measure on the product space $\prod_{x\in\Lambda^*}B_q(0,1)$.
Next we give a precise meaning to the approximation
$\mu_{\beta,\gamma,\Lambda}\approx\exp[-\beta\gamma^{-d}{\cal F}(m|m^c)]$ stated in the
Introduction, by establishing the analogue of [AlBeCaPr,Lemma 3.2] in case of continuous symmetry,
improving also [BuPi, Lemma 3.1]~:
\medskip
\noindent{\bf Theorem  c.4}: With the notations above, there are constants $C_1, C_2>0$ such that
for any coarse-grained configuration $m$ on $\Lambda$, we have~:
$$\eqalign{
-g(m)&-(L\gamma^{-1})^d\bigl(C_2\beta\gamma^\delta\|\nabla J\|_1+C_1\gamma^{(1-\delta)d}\log
\gamma^{-1}\bigr)\cr \leq\log & \
\mu_{\beta,\gamma,\Lambda}(d\sigma_\gamma;m|\sigma_\gamma^c)+\beta\gamma^{-d}{\cal F}(m|m^c)\cr
&\leq g(m)+\inf_{\widehat e\in{\bf S^1}}{\cal F}(\widehat m_\beta|m^c)
+(L\gamma^{-1})^d\bigl(C_2\beta\gamma^\delta\|\nabla J\|_1+C_1\gamma^{(1-\delta)d}\log
\gamma^{-1}\bigr)\cr }\leqno(c.35)$$ where $g(m)=\log\prod_{x\in\widetilde \Lambda/\widetilde
C^{(n_\gamma)}}\bigl(1-|m(x)|\bigr)^{-1/2}$.
\smallskip
\noindent{\it Proof}: First we look for a lower bound on
$Z_{\beta,\gamma,\Lambda}(\sigma_\gamma^c)$. Using (c.30) we get~:
$$\exp [-\beta\gamma^{-d}H(\sigma_\gamma|\sigma^c_\gamma)]\geq
\exp [-\beta\gamma^{-d}H(\pi_\gamma\sigma_\gamma|\sigma^c_\gamma)] \exp  [-C_2\beta\gamma^{-d}L^d
\gamma^\delta\|\nabla J\|_1]$$ (and similarly for the upper bound). Multiply these relations by
$\delta(\pi_\gamma\sigma_\gamma(i)-m)$, integrate with respect to
$\prod_{i\in\widetilde\Lambda}\nu\bigl(d\sigma_{\widetilde\Lambda}(i)\bigr)$ over $\Omega_0$ and
use (c.25), we get
$$\eqalign{
&\exp[-\beta\gamma^{-d}(H(m|m^c)+{1\over\beta}I(m,\Lambda))]\exp-\psi_\gamma(m)\cr
&\leq\int_{\Omega_0}\prod_{i\in\widetilde\Lambda}\nu\bigl(d\sigma_{\widetilde\Lambda}(i)\bigr) \exp
[-\beta\gamma^{-d}H(\sigma_\gamma|\sigma^c_\gamma)] \delta(\pi_\gamma\sigma_\gamma(i)-m)\cr
&\leq\exp[-\beta\gamma^{-d}(H(m|m^c)+{1\over\beta}I(m,\Lambda))]\exp\psi_\gamma(m)\cr
}\leqno(c.36)$$ where
$$\psi_\gamma(m)=(L\gamma^{-1})^d\bigl(C_2\beta\gamma^\delta\|\nabla J\|_1+C_1\gamma^{(1-\delta)d}\log \gamma^{-1}\bigr)
+g(m) \leqno(c.37)$$ Now we estimate the contribution of a neighborhood of $\widehat m_\beta$ to
the partition function. So let $0\leq \chi\leq 1$ be a smooth positive cutoff equal to 1 near
$m=\widehat m_\beta$, multiply (c.36) by $\chi(m)$ and integrate over $m$ with respect to the
product measure $\prod_{x\in\widetilde\Lambda/\widetilde C^{(n_\gamma)}}dm(x)$, then make use of
(c.9), (c.32) and (c.34), we get
$$\eqalign{
Z_{\beta,\gamma,\Lambda}(\sigma_\gamma^c)&\geq\exp[-\|\psi_\gamma\|_\chi]
\exp[-\beta\gamma^{-d}(H(\widehat m_\beta|m^c)+{1\over\beta}I(\widehat m_\beta,\Lambda))]\cr
&\times\int dm\chi(m) \exp[-\beta\gamma^{-d}({\cal F}(m|m^c)-{\cal F}(\widehat m_\beta|m^c)] }$$
where $\|\psi_\gamma\|_\chi=\sup _{m\in \supp \chi}\psi_\gamma(m)<\infty$. Choose supp $\chi$ so
small that $|m-\widehat m_\beta|\leq\gamma^{(1-\delta)d}$ on supp $\chi$. Using (c.31), the
normalisation of $J$ and Taylor expansion of $f_\beta$ around $\widehat m_\beta$, we get $|{\cal
F}(m|m^c)-{\cal F}(\widehat m_\beta|m^c)|\leq C_3 L^d \gamma^{(1-\delta)d}$ on supp $\chi$, so~:
$$Z_{\beta,\gamma,\Lambda}(\sigma_\gamma^c)\geq\exp[-\bigl(\|\psi_\gamma\|_\chi+C_3 (L\gamma^{-1})^d\beta \gamma^{(1-\delta)d}
\bigr)] \exp[-\beta\gamma^{-d}(H(\widehat m_\beta|m^c)+{1\over\beta}I(\widehat
m_\beta,\Lambda))]\leqno(c.38)$$ Inserting this in (c.33) and (c.36) we find
$$
\log\mu_{\beta,\gamma,\Lambda}(d\sigma_\gamma;m|\sigma_\gamma^c)\leq -\beta\gamma^{-d}({\cal
F}(m|m^c)-{\cal F} (\widehat m_\beta|m^c))+\psi_\gamma(m)+\|\psi_\gamma\|_\chi+C_3
(L\gamma^{-1})^d\beta \gamma^{(1-\delta)d} \leqno(c.39)$$ For the upper bound on
$Z_{\beta,\gamma,\Lambda}(\sigma_\gamma^c)$, we use (c.34) and (c.36) to write
$$\eqalign{
&Z_{\beta,\gamma,\Lambda}(\sigma_\gamma^c)=\int dm\int_{\Omega_0} \exp
\bigl[-\beta\gamma^{-d}H_\gamma(\sigma_\gamma|\sigma_\gamma^c)\bigr]
\prod_{i\in\widetilde\Lambda}\nu\bigl(d\sigma_{\widetilde\Lambda}(i)\bigr)\delta(\pi_\gamma\sigma_\gamma(i)-m)\cr
&\leq\int dm\exp[-\beta\gamma^{-d}(H(m|m^c)+{1\over\beta}I(m,\Lambda))]\exp\psi_\gamma(m)\cr }$$ By
(c.32) and inequality ${\cal F}(m|m^c)\geq 0$, we have
$$\eqalign{
Z_{\beta,\gamma,\Lambda}&(\sigma_\gamma^c)\leq\exp[-\beta\gamma^{-d}\bigl( H(\widehat
m_\beta|m^c)+{1\over\beta} I(\widehat m_\beta,\Lambda))-{\cal F}(\widehat m_\beta|m^c)\bigr)]\cr
&\times\int dm\exp\psi_\gamma(m)\cr }$$ se we are left to estimate $\int dm\exp\psi_\gamma(m)$, the
integral running over the product space, and $\psi_\gamma(m)$ as in (c.37). Since
$\int_0^1(1-\rho)^{-1/2}\rho d\rho<\infty$  we find
$$\eqalign{
Z_{\beta,\gamma,\Lambda}&(\sigma_\gamma^c)\leq \exp[-\beta\gamma^{-d}\bigl( H(\widehat
m_\beta|m^c)+{1\over\beta} I(\widehat m_\beta,\Lambda)-{\cal F}(\widehat m_\beta|m^c)\bigr)]\cr
&\times\exp[(L\gamma^{-1})^d\bigl(C_2\beta\gamma^\delta\|\nabla J\|_1+C_1\gamma^{(1-\delta)d}\log
\gamma^{-1} +C_4\gamma^{(1-\delta)d}\bigr)]\cr }$$ Inserting this and the first inequality (c.36)
in (c.33), we find, absorbing the $C_4$-remainder term into the $C_1$-remainder term, and using
$$\eqalign{
\log\mu_{\beta,\gamma,\Lambda}&(d\sigma_\gamma;m|\sigma_\gamma^c)\geq -g(m)-\beta\gamma^{-d}{\cal
F}(m|m^c)\cr &-(L\gamma^{-1})^d\bigl(2C_2\beta\gamma^\delta\|\nabla J\|_1
+2C_1\gamma^{(1-\delta)d}\log \gamma^{-1}\bigr)\cr }\leqno(c.43)$$ Putting (c.43) together with
(c.39) with new constants $C_1, C_2$ gives the Theorem. $\clubsuit$
\medskip
Of course,these estimates break down when $|m(x)|$ gets close to 1 for some $x\in\Lambda$, which
reflects the fact that the entropy density $I(m)$ is singular near $|m|=1$. It is shown in
[BuPi,Theorem 2.2] that $\nu_N(\{|m|>1-\rho\})$ decays exponentially fast as $N\to\infty$ when
$\rho>0$ is small enough.

\bigskip
\centerline{\bf References}
\medskip
\noindent [AlBeCaPr] G.Alberti, G.Bellettini, M.Cassandro, E.Presutti. Surface tension in Ising
systems with Kac potentials. J. Stat. Phys. Vol.82, (3 and 4) (1996) p.743-795.

\noindent [AlBe] G.Alberti, G.Bellettini. A nonlocal anisotropic model for phase transitions. Math.
Ann. 310, (1998) p.527-560.

\noindent [Al] G.Alberti. Some remarks about a notion of rearrangements. Ann. Scuola Norm. Sup.
Pisa Cl. Sci. (4), Vol. XXIX (2000), p.457-472.

\noindent [BetBrHe] F.Bethuel, H.Brezis, F.Helein. Ginzburg-Landau vortices, Birkh\"auser, Basel,
1994.

\noindent [BuPi] P.Butt\`a, P.Picco. Large-deviation principle for one-dimensional vector spin
models with Kac potentials. J. Stat. Phys. Vol.92, (1 and 2) (1998) p.101-150.

\noindent [Da] B.Dacorogna. Direct methods in the Calculus of Variations. Springer, Berlin,
Heidelberg 1989.

\noindent [DeM] A.DeMasi. Spins systems with long range interactions. Progress in Probability,
Birkh\"auser, Vol 54 (2003), p.25-81.

\noindent [DeMOrPrTr] A.DeMasi, E.Orlandi, E.Presutti, L.Triolo. {\bf 1} Motion by curvature by
scaling nonlocal evolution equations. J. Stat. Phys. 73 (1993), p.543-570. {\bf 2} Glauber
evolution with Kac potentials I. Nonlinearity 7 (1994), p.663-696. {\bf 3} Stability of the
interface in a model of phase separation, Proc. Royal Soc. Edinburgh 124-A (1994), p.1013-1022.
{\bf 4} Uniqueness and global stability of the instanton in non-local evolution equations.
Rendiconti di Mat., Serie VII, 14, (1994), p.693-723.

\noindent [El-BoRo] H.El-Bouanani, M.Rouleux. Vortices and magnetization in Kac's model. J. Stat.
Phys. 2007.

\noindent [FiMc-L] P.Fife, J.B.McLoad. The approach of solutions of nonlinear diffusion equations
to travelling front solutions. Arch. Rat. Mech. Anal. 75 (1977) p.335-361.

\noindent [FoGa] I.Fonseca, W.Gangbo. Degree theory in analysis and applications. Clarendon Press,
Oxford, 1995.

\noindent [FrTo] S.Franz, F.L.Toninelli. The Kac limit for finite-range spin systems.

\noindent [GiVe] J.Ginibre, G.Velo.  The Cauchy problem in  local spaces for the complex
Ginzburg-Landau equation {\bf  I}. Compactness methods. Physica D 95, 1996 p.191-228. {\bf II}.
Contraction methods. Commun. Math. Phys. 187 (1997), p.45-79.

\noindent [L] N.Lebedev. Special Functions and their Applications. Dover, N.Y., 1982.

\noindent [LePe] J.Lebowitz, O.Penrose, J.Math. Physics. Vol.7 (1) (1966) p.98-113

\noindent [LiLo] E.Lieb, M.Loss. Symmetry of the Ginzburg-Landau minimizer in a disc. Proceedings
Journ\'ees Equations d\'eriv\'ees partielles, Saint-Jean-de-Monts (1995), p.1-12.

\noindent [Mi] P. Mironescu. On the stability of radial solutions of the Ginzburg-Landau equation.
J. Funct. Anal. Vol.130 (1995) p.334-344.

\noindent [Mo] E.Mourre. Absence of singular continuous spectrum for certain self-adjoint
operators. Commun. Math. Phys. 78, (1981), p.391-408.

\noindent [OvSi] Y.Ovchinnikov, I.M.Sigal. {\bf 1} Ginzburg-Landau Equation I. Static vortices. CRM
Proceed. Vol.12, (1997) p.199-220, {\bf 2} The energy of Ginzburg-Landau vortices, European J. of
Applied Mathematics 13, (2002) p.153-178.

\noindent [Pr] E.Presutti. From statistical mechanics towards continuum mechanics. Preprint
M.Planck Institut, Leipzig, 1999

\noindent [ReSi] M.Reed, B.Simon. Modern methods of mathematical physics, Vol IV, Analysis  of
operators. Academic Press, 1978.

\noindent [Se] S.Serfaty. Vorticit\'e dans les \'equations de Ginzburg-Landau de la
supraconductivit\'e. S\'eminaire EDP Ecole Polytechnique, Expos\'e no VI, 1999-2000.

\noindent [Si] Ya.G.Sina\"i. Theory of Phase Transitions~: Rigorous Results. Pergamon, Oxford,
1982.

\noindent [Sj]  J.Sj\"ostrand. Singularit\'es  analytiques microlocales. Ast\'erisque 95. Soc.
Math. France, 1982.

\noindent [Ti] E.C.Titschmarsch. Hankel transforms. Proc. Cambridge Phil. Soc. 21, p.463-473, 1923.

\noindent [Wa] G.N.Watson. A Treatise on the Theory of Bessel functions. Cambridge  University
Press, London, 1944.

\bye